# Stationary and time dependent correlations in polariton condensates

Paolo Schwendimann and Antonio Quattropani

Institute of Theoretical Physics, Ecole Polytechnique Fédérale de Lausanne, CH 1015

Lausanne-EPFL, Switzerland

and

Davide Sarchi

CNR-INFM BEC Center and

Dipartimento di Fisica, Università di Trento, I-38050 Povo, Trento, Italy

The statistics of the condensed polaritons is described in terms of the Wigner function. In the framework of the truncated Wigner method, the Wigner function obeys a Fokker-Planck equation, which is solved analytically. The second order correlations in the stationary state are in excellent agreement with those obtained from the numerical solution of the master equation and show a qualitative and, well above threshold, also quantitative agreement with recent experiments. Furthermore, the contributions of the different noise effects that influence the polariton ground state statistics are explicitly defined. Exploiting the equivalence between Fokker-Planck and Langevin descriptions of stochastic processes, the time dependent correlations of the polaritons close to the stationary state are derived. Explicit expressions for the linewidth and for the relaxation rate of the polariton intensity are obtained, whose values are of the same order of magnitude of the experimental data. Finally, the limit of validity of the truncated Wigner method in the present model is discussed.

71.35.Lk, 71.36.+c, 71.55.Gs

## I. INTRODUCTION

Long before the demonstration of Bose-Einstein condensation in atomic systems [1,2], the question has been raised whereas analogous effects could be detected in solid-state systems. The candidates for such a condensation effect are excitations, which show a boson-like behavior. Examples of such excitations are the exciton and the exciton-polariton in a semiconductor microcavity. The exciton-polariton can be assumed to be a boson for not to high excitation intensities and it has a very small mass. This last characteristic indicates that condensation may take place at reasonably high temperatures. Indeed, the insurgence of a strong emission into the polariton state with the wave vector k = 0 was observed in a CdTe sample [3] by exciting a semiconductor quantum well into the conduction band. The insurgence of a macroscopically populated state at k = 0 analogous to a laser state was demonstrated in a GaAs sample [4]. Polariton Bose-Einstein condensation was finally demonstrated in a CdTe sample [5]. In the following years signatures of polariton condensation have been found in different samples including GaAs [6], and GaN [7] and several properties of the condensed state have been extensively studied [8]. Notice that, in contrast to a gas where the condensation happens in thermal equilibrium, polariton condensation is a non-equilibrium effect resulting from the interplay between the external pump and the finite lifetime of the polariton as well as the cavity losses.

In experiments on polariton condensation one measures quantities that are related to the photonic component of the polariton, which is emitted from the microcavity. Therefore, it is expected that the techniques developed in quantum optics shall be successfully applied in the investigation of the condensed state. In particular, the photon statistics of the emitted field offers a powerful tool for investigating some characteristics of the polariton condensate through the determination of its coherence properties. For example, in the ideal case one expects the polariton condensate to exhibit a high level of coherence. Deviations from this ideal state indicate that some peculiar scattering processes between condensed and non-condensed polaritons are present. The relevance of these processes to the coherence of the polaritons as a function of the material and excitation pump characteristics is studied both theoretically and experimentally through the polariton correlation functions. Indeed, the first measurement of the second order polariton correlation was performed in 2002 in a GaAs sample [9]. The correlation shows a slow decay from its incoherent value of two as a function of the pump above the threshold but full coherence is not achieved. After the demonstration of polariton condensation [5] the stationary second order correlation was measured [10] in a CdTe microcavity. Just above threshold full second order coherence was achieved, but for larger values of the pump intensity the full second order coherence was lost. Further measurements in a GaAs microcavity [11] show that well above threshold the full coherence is not found. Furthermore, the third order correlation shows a behavior that is quite different from that expected when full coherence is present.

From a theoretical viewpoint, the peculiar behavior of the second order correlation observed in the experiments is a consequence of the parametric scattering between polaritons in the condensate and non-condensed polaritons having opposite wave vectors. These scattering processes act as a noise source in the condensate thus affecting its coherence properties. In fact, in the first theoretical calculations of the statistics of the polariton condensate [12], these processes are not considered and the polariton statistics is that of the conventional laser. The relevance of the parametric processes to the coherence of the condensate has been discussed in Refs. [13, 14] in the framework of master equation formalism and exploiting the polariton bottleneck when modeling the evolution of the non-condensed polaritons. In [13] the bottleneck polaritons were considered to act as a reservoir whose temperature was fixed and whose population density was determined by the external pump intensity. Although the effects of the parametric scattering were included in this approach the statistics of the polaritons was not different from that of a laser due to an underestimation of the parametric scattering rates in the framework of the chosen reservoir model. In [14] the non-condensed polariton dynamics was modeled following the lines of Porras et al. [15] where the bottleneck polariton temperature and density are self-consistently determined and the dynamics of the polaritons not pertaining neither to the bottleneck region nor to the condensate is accounted for. In this case, the polariton correlations show a behavior that qualitatively reproduces the one found in the experiments, but this time the effects of the parametric scattering are overestimated as shown in [16]. In this last paper the reservoir model is replaced by the full Boltzmann dynamics of the polaritons including the polariton-phonon interaction. The result although qualitative, indicates that the behavior of the polariton correlations may be well understood in this framework. We also mention that calculations of the polariton correlations are discussed in [17] on lines similar to [16] and in [18] in a different framework starting from the Gross-Pitaiewskii equation. However, the correlation calculated in these papers don't compare well with the experiments.

The motivation of the present paper lies in the very recent measurements of the correlations of the polaritons in the ground state including third order correlations [11]. Like in the previous papers [13, 14], we consider a quantum well embedded in a semiconductor microcavity and excited into the conduction band by a laser pump in the cw regime. We also recall that the dynamics of the polaritons is determined by the scattering processes involving polariton population as well as by the parametric scattering processes mentioned above. We shall mainly be interested in the stationary behavior of the polariton system. In order to understand the experimental results in [11], in the spirit of [16] we have chosen a different point of view, which allows to better understand the role of the parametric scattering and furthermore results in a major speeding up of the calculations. Instead of solving the master equation as in [13, 14] we transform it into a differential equation for the Wigner function and solve this equation within the so-called truncated Wigner method [19]. In this approximation, which consists in neglecting third order derivatives in the original equation, the Wigner function obeys a Fokker-Planck equation. The advantages of our approach are threefold:

(i) As we shall see in the following, we obtain an analytical expression for the approximated stationary Wigner function from which all moments are rapidly calculated. These approximated moments show an astonishing good agreement with the results from the master equation whose solution requires long calculation times. Furthermore, as it is shown in [11] there is a satisfactory agreement between the theoretical results calculated in this framework and the experimental results.

(ii) We exploit the relation between the Fokker-Planck and the Langevin description of stochastic processes in order to obtain a Langevin description of the polariton dynamics, from which the explicit expressions of the linewidth and of relaxation rate of the second order correlation close to the stationary state are obtained. We show that the linewidth is related to the diffusion rate of the phase of the polaritons in the ground state. This rate results from the interplay between the noise contributions originating both in the scattering processes involving the polariton populations and in the parametric scattering processes. Above threshold we show that the latter processes mainly determine the behavior of the linewidth as a function of the pump. The values obtained for the linewidth are comparable with the ones obtained from the experiments. The relaxation rate of the second order correlation mainly depends on the gain and saturation effects in the condensate. These are in part compensated above threshold by the noise originating in the parametric scattering. Also in this case, the calculated rates are comparable with the experimental ones.

(iii) We have obtained the analytic expression for the exact stationary Wigner function thus allowing us to discuss quantitatively the validity of the truncated Wigner method in this particular case. We show that the approximate and exact Wigner functions have qualitatively the same behavior as functions of the polariton amplitude above and around threshold but strongly differ below threshold. This behavior reflects in the correlations calculated with the approximate or exact solution respectively. We find that above and around threshold the correlations calculated with both the approximate and the exact solution show an irrelevant difference when comparing the first ten moments. However, below threshold there are differences between the moments as functions of the external pump that are more pronounced with growing order of the correlations consistently with the difference between the approximated and the exact solutions. Therefore the validity of the truncated Wigner method is not assured in this regime.

In the spirit of the above considerations, we have organized the paper as follows. In Section II we derive the equation for the Wigner function and present its approximate solutions. A discussion of its moments follows in Section III. In Section IV we derive the explicit expression for the linewidth and the relaxation rate from a Langevin approach, and finally in Section V we derive the exact Wigner function and discuss the limits of validity of the approximate one. Details on the derivation of the master equation, which is used as the starting point in our approach, are given in the Appendix.

## II. APPROXIMATE QUASI-PROBABILITY DISTRIBUTION

In this Section we derive an approximate analytical expression for a quasi-probability distribution of the polariton condensate. From this distribution the statistical properties of the condensate are calculated. The quasi-probability distribution is obtained by exploiting the relation between the density operator and its c-number representations introduced by Glauber [20-23] in quantum optics. Firstly, we derive a master equation, which, contrary to those of Refs.[13, 14], doesn't relay on the introduction of a polariton reservoir. Furthermore, in deriving this master equation, we introduce the phonon-polariton interaction in order to account for the relaxation of the polaritons from the excited state created by an external non-resonant pump. Here, we only mention the most relevant steps of its derivation, whose details are given in the Appendix. We start with some considerations on the scattering processes between polaritons described by the polariton-polariton Hamiltonian [24]

$$H = H_0 + H_{int} = \sum_{\mathbf{k}} \hbar \omega_{\mathbf{k}} P_{\mathbf{k}}^+ P_{\mathbf{k}} + \sum_{\mathbf{k},\mathbf{k}',\mathbf{q}} W_{\mathbf{k},\mathbf{k}',\mathbf{q}} P_{\mathbf{k}+\mathbf{q}}^+ P_{\mathbf{k}'-\mathbf{q}}^+ P_{\mathbf{k}} P_{\mathbf{k}'} \qquad (2.1)$$

The explicit expressions of the coupling constants $W_{\mathbf{k},\mathbf{k}',\mathbf{q}}$ in (2.1) are given in the Appendix. The interaction part of (2.1) contains the contribution of two different scattering processes involving polariton pairs and satisfying momentum conservation. The first one consists of the scattering processes between polariton pairs $P_{\mathbf{k}+\mathbf{q}}^+ P_{\mathbf{k}'-\mathbf{q}}^+ P_{\mathbf{k}} P_{\mathbf{k}'}$ with two different wave vectors ($\mathbf{k} \neq \mathbf{k}'$) both in the initial and in the final state. They are responsible for the polariton gain and relaxation processes. The second one consists of the parametric scattering of polariton pairs with $\mathbf{k} = \mathbf{k}'$ (the source) into polaritons pairs with wave vectors $\mathbf{k}+\mathbf{q}$ (the idler) and $\mathbf{k}-\mathbf{q}$ (the signal) respectively, described by terms of the form $P_{\mathbf{k}+\mathbf{q}}^+ P_{\mathbf{k}-\mathbf{q}}^+ P_{\mathbf{k}} P_{\mathbf{k}}$. As we shall see in the following, the parametric scattering plays an important role in the characterization of the noise in the polariton condensate. In order to show how these scattering processes contribute in the dynamics of the density operator of the polariton condensate, we start from le Liouville-von Neumann equation for the density operator of the polariton system described by (2.1)

$$\hbar \frac{d\rho}{dt} = -i[H, \rho]. \tag{2.2}$$

We take the trace of (2.2) over all wave vectors with $\mathbf{k} \neq 0$ obtaining

$$\hbar \frac{d\rho_0(t)}{dt} = -i\left\{\left[P_0^+, \sum_{\mathbf{k},\mathbf{k}'\neq 0} W_{\mathbf{k},\mathbf{k}'0} Tr_{\mathbf{k},\mathbf{k}'\neq 0}\left(P_{\mathbf{k}+\mathbf{k}'}^+ P_{\mathbf{k}-\mathbf{k}'}^+ P_{\mathbf{k}'}\rho(t)\right)\right] + h.c.\right\} -$$
$$i\left\{\left[P_0^2, \sum_{\mathbf{k}\neq 0} W_{\mathbf{k},-\mathbf{k},0} Tr_{\mathbf{k},-\mathbf{k}\neq 0}\left(P_{\mathbf{k}}^+ P_{-\mathbf{k}}^+ \rho(t)\right)\right] + h.c.\right\} -$$
$$i\hbar\omega_0\left[P_0^+ P_0, \rho_0(t)\right] - iW_{0,0,0}[P_0^+ P_0^+ P_0 P_0, \rho_0(t)] \tag{2.3}$$

Notice that in (2.2) dissipation is not yet included. The commutators on the r.h.s. of (2.3) represent the separate contributions of the two different scattering processes quoted above to the evolution of $\rho_0$. In order to obtain the master equation from (2.3) one has to express explicitly the operators $Tr_{\mathbf{k},\mathbf{k}'\neq 0}\left(P_{\mathbf{k}+\mathbf{k}'}^+ P_{\mathbf{k}-\mathbf{k}'}^+ P_{\mathbf{k}'}\rho(t)\right)$ and $Tr_{\mathbf{k}\neq 0}\left(P_{\mathbf{k}}^+ P_{-\mathbf{k}}\rho(t)\right)$ in (2.3) in terms of the operators $P_0^+, P_0, \rho_0(t)$. The expressions for these contributions are derived in the Appendix starting from the full Liouville-von Neumann equation that contains besides the polariton-polariton Hamiltonian the polariton-phonon interaction Hamiltonian and a dissipative term describing the cavity losses. The main approximations introduced in this derivation are the Markov approximation and the following factorization Ansatz

$$Tr_{\{\mathbf{k}\}\neq 0}\left(P_{\mathbf{k}}^{+n} P_{\mathbf{k}}^{n}\rho\right) = \rho_0 \left(<P_{\mathbf{k}}^+ P_{\mathbf{k}}>\right)^n \tag{2.4a}$$
$$Tr_{\{\mathbf{k}\}\neq 0}\left(P_{\mathbf{k}}^{+m} P_{\mathbf{k}}^{n}\rho\right) = 0 \qquad m \neq n \tag{2.4b}$$

The master equation then reads

$$\hbar \frac{d}{dt}\rho_0(t) = -i\hbar\omega_0\left[P_0^+P_0, \rho_0(t)\right] - iW_{0,0,0}[P_0^+P_0^+P_0P_0, \rho_0(t)] + \Lambda_0\rho_0(t)) +$$
$$\Gamma_0\left(\left[P_0\rho_0(t), P_0^+\right] + h.c.\right) + \Delta_0\left(\left[P_0^+\rho_0(t), P_0\right] + h.c.\right)$$
$$\Gamma_1\left(\left[P_0^2\rho_0(t), P_0^{+2}\right] + h.c.\right) + \Delta_1\left(\left[P_0^{+2}\rho_0(t), P_0^2\right] + h.c.\right) \tag{2.5}$$

with the definitions

$$\Gamma_0 = 2\operatorname{Re}\sum_{\mathbf{k},\mathbf{k}'\neq 0,} G_{\mathbf{k},\mathbf{k}',0}W_{\mathbf{k},\mathbf{k}',0}^2\left[\left\langle P_{\mathbf{k}+\mathbf{k}'}^+P_{\mathbf{k}+\mathbf{k}'}\right\rangle\left(\left\langle P_{\mathbf{k}}^+P_{\mathbf{k}}\right\rangle+1\right)\left(\left\langle P_{\mathbf{k}'}^+P_{\mathbf{k}'}\right\rangle+1\right)\right] \tag{2.6a}$$

$$\Delta_0 = 2\operatorname{Re}\sum_{\mathbf{k},\mathbf{k}'\neq 0,} G_{\mathbf{k},\mathbf{k}',0}W_{\mathbf{k},\mathbf{k}',0}^2\left[\left(\left\langle P_{\mathbf{k}+\mathbf{k}'}^+P_{\mathbf{k}+\mathbf{k}'}\right\rangle+1\right)\left\langle P_{\mathbf{k}}^+P_{\mathbf{k}}\right\rangle\left\langle P_{\mathbf{k}'}^+P_{\mathbf{k}'}\right\rangle\right] \tag{2.6b}$$

$$\Gamma_1 = 2\operatorname{Re}\sum_{\mathbf{k}\neq 0} G_{\mathbf{k},0}W_{\mathbf{k},-\mathbf{k},0}^2\left(\left(<P_{\mathbf{k}}^+P_{\mathbf{k}}>+1\right)\left(<P_{-\mathbf{k}}^+P_{-\mathbf{k}}>+1\right)\right) \tag{2.6c}$$

$$\Delta_1 = 2\operatorname{Re}\sum_{\mathbf{k}\neq 0} G_{\mathbf{k},0}W_{\mathbf{k},-\mathbf{k},0}^2\left(<P_{\mathbf{k}}^+P_{\mathbf{k}}><P_{-\mathbf{k}}^+P_{-\mathbf{k}}>\right) \tag{2.6d}$$

The quantities $G_{\mathbf{k},\mathbf{k}',0}$, $G_{\mathbf{k},0}$ and $\Lambda_0$ are defined in the Appendix. The populations $\left\langle P_{\mathbf{k}}^+P_{\mathbf{k}}\right\rangle$ that appear in (2.6) are calculated by solving a coupled system of non-linear time-dependent differential equations that are derived in the Appendix in the same framework as the master equation (2.5). As a consequence, the populations in (2.6) are time dependent quantities. Since in the following we shall be concerned with the stationary behavior of the system, we don't show this time dependence explicitly in (2.5) and (2.6). In (2.5), the terms with the dissipation rate $\Gamma_0$ and the injection rate $\Delta_0$ respectively originate in the first commutator in (2.3) and are related to the non-parametric pair scattering, whereas the terms with the dissipation rate $\Gamma_1$ and the injection rate $\Delta_1$ respectively originate in the second commutator in (2.3) and are related to the parametric scattering.

The equations (2.5) and (2.6) are formally identical with the master equations presented in [13, 14] but the expectation values for the polariton populations with wave vector $\mathbf{k} \neq 0$ are calculated in a different framework. In Ref. [13] these expectation values were calculated by assuming that the polaritons with $\mathbf{k} \neq 0$ belong to a reservoir whose occupation is determined by the pump intensity. This approximation leads to an underestimation of the rates (2.6c) and (2.6d). In Ref. [14] the expectation values in (2.6c) and (2.6d) have been calculated by generalizing the approach of [15]. In this approach the dynamics of the polaritons below the bottleneck is described by Boltzmann-like equations, whereas the polaritons in the bottleneck are described by a self consistently defined equilibrium state. This approach overestimates the effect of the terms (2.6c) and (2.6d) as it is shown in [16]. The full dynamics of the polaritons with $\mathbf{k} \neq 0$ in the calculation of the polariton statistics has also been considered in [12, 17] in the framework of the polariton rate equation scheme [25], that coincides with the one presented here up to the terms resulting from the parametric scattering. The master equation considered in [12] however doesn't contain the contribution of the parametric scattering effects thus leading to an oversimplified description of the polariton statistics.

The statistics of the polariton condensate in the stationary state has been obtained from the numerical solution of the diagonal part of (2.5) whereas its off-diagonal part is shown to vanish in the stationary regime [13]. However, solving numerically (2.5) requires very long calculation times depending on the material parameters. We gain more efficiency and a better insight into the physics described by (2.5) by transforming it into a differential equation for a quasi-probability function. We have chosen to work with the Wigner function as it is currently done when discussing Bose-Einstein condensation. The Wigner function is a quasi-probability distribution that is defined as the Fourier transform of the characteristic function of the system. The Wigner function for the density operator of the ground state polaritons is defined as

$$W(\alpha,\alpha^*) = \int d^2\beta \exp(\beta^*\alpha - \beta\alpha^*)\chi(\beta,\beta^*) \qquad (2.7a)$$

with the characteristic function

$$\chi(\beta,\beta^*) = Tr\left\{\exp\left[\beta P_0^+ - \beta^* P_0\right]\rho_0\right\}, \qquad (2.7b)$$

and

$$\left[P_0, P_0^+\right] = 1 \qquad (2.7c)$$

We refer to the standard literature [26] for its properties and for examples of the transformation of an operator master equation into a c-number equation for the Wigner function. Using these standard techniques we get from (2.5) the partial differential equation

$$\begin{aligned}
\hbar\frac{\partial W(\alpha,\alpha^*)}{\partial t} = &-i\hbar\omega_0\left(\frac{\partial}{\partial\alpha}\alpha - \frac{\partial}{\partial\alpha^*}\alpha^*\right)W(\alpha,\alpha^*) - \\
&iW_{0,0,0}\left(\frac{\partial}{\partial\alpha}|\alpha|^2\alpha - \frac{\partial}{\partial\alpha^*}|\alpha|^2\alpha^*\right)W(\alpha,\alpha^*) - \\
&\frac{iW_{0,0,0}}{2}\frac{\partial^2}{\partial\alpha\partial\alpha^*}\left(\frac{\partial}{\partial\alpha}\alpha - \frac{\partial}{\partial\alpha^*}\alpha^*\right)W(\alpha,\alpha^*) + \\
&\left[\Gamma_0 + \gamma_0 - \Delta_0 - 2(\Gamma_1 + \Delta_1)\right]\left(\frac{\partial}{\partial\alpha}\alpha + \frac{\partial}{\partial\alpha^*}\alpha^*\right)W(\alpha,\alpha^*) + \\
&2(\Gamma_1 - \Delta_1)\left(\frac{\partial}{\partial\alpha}|\alpha|^2\alpha + \frac{\partial}{\partial\alpha^*}|\alpha|^2\alpha^*\right)W(\alpha,\alpha^*) + \\
&\frac{\partial^2}{\partial\alpha\partial\alpha^*}(\Gamma_0 + \gamma_0 + \Delta_0)W(\alpha,\alpha^*) + 4(\Gamma_1 + \Delta_1)\frac{\partial^2}{\partial\alpha\partial\alpha^*}|\alpha|^2 W(\alpha,\alpha^*) + \\
&\frac{(\Gamma_1 - \Delta_1)}{2}\left(\frac{\partial^2}{\partial\alpha^*}\alpha\frac{\partial}{\partial\alpha^2} + \frac{\partial}{\partial\alpha}\alpha^*\frac{\partial}{\partial\alpha^{*2}}\right)W(\alpha,\alpha^*)
\end{aligned} \qquad (2.8)$$

The equation (2.8) is a third order partial differential equation. As it is often done in discussing the statistical properties of Bose-Einstein condensates [19] we simplify equation (2.8) by neglecting the contribution of the third order derivatives. . As we shall show in the following this approximation is a very good one when compared to the numerical results from the master equation (2.5) for most values of the external pump both above and around threshold. The approximate equation reads

$$\hbar \frac{\partial W(\alpha,\alpha^*)}{\partial t} = -i\hbar\omega_0 \left( \frac{\partial}{\partial \alpha}\alpha - \frac{\partial}{\partial \alpha^*}\alpha^* \right) W(\alpha,\alpha^*) -$$

$$iW_{0,0,0} \left( \frac{\partial}{\partial \alpha}|\alpha|^2 \alpha - \frac{\partial}{\partial \alpha^*}|\alpha|^2 \alpha^* \right) W(\alpha,\alpha^*) +$$

$$\left[ \Gamma_0 + \gamma_0 - \Delta_0 - 2(\Gamma_1 + \Delta_1) \right] \left( \frac{\partial}{\partial \alpha}\alpha + \frac{\partial}{\partial \alpha^*}\alpha^* \right) W(\alpha,\alpha^*) +$$

$$2(\Gamma_1 - \Delta_1) \left( \frac{\partial}{\partial \alpha}|\alpha|^2 \alpha + \frac{\partial}{\partial \alpha^*}|\alpha|^2 \alpha^* \right) W(\alpha,\alpha^*) +$$

$$\frac{\partial^2}{\partial \alpha \partial \alpha^*}(\Gamma_0 + \gamma_0 + \Delta_0) W(\alpha,\alpha^*) + 4(\Gamma_1 + \Delta_1) \frac{\partial^2}{\partial \alpha \partial \alpha^*}|\alpha|^2 W(\alpha,\alpha^*) \qquad (2.9a)$$

We recognize in (2.9) a Fokker-Planck equation in which the real part of the drift term is

$$\left[ \Gamma_0 + \gamma_0 - \Delta_0 - 2(\Gamma_1 + \Delta_1) \right]\alpha + 2(\Gamma_1 - \Delta_1)|\alpha|^2 \alpha \qquad (2.9b)$$

and the diffusion term is

$$(\Gamma_0 + \gamma_0 + \Delta_0) + 4(\Gamma_1 + \Delta_1)|\alpha|^2 \qquad (2.9c)$$

Both quantities depend on the polariton amplitude. The real part of the drift consists of firstly a term linear in α that contains the quantity $\Gamma_0 + \gamma_0 - \Delta_0$, representing the gain i.e. the difference between the losses characterized by the dissipation rate $\Gamma_0 + \gamma_0$ and the injection rate $\Delta_0$ into the polariton ground state. This quantity changes its sign in function of the intensity of the external pump as shown in [13, 14]. Secondly, it contains a term non-linear in α inducing saturation effects, which depend only on the parametric scattering rate and stabilize the solution above threshold. The diffusion coefficient too, depends both on the non-parametric and on the parametric scattering. In particular, the field-dependent part of the diffusion coefficient depends the parametric scattering rates only. In the following we call this term "parametric diffusion coefficient". This term plays a central role in determining the statistical properties of the polariton system in the ground state. In fact, by neglecting the parametric diffusion, equation (2.9a) becomes a Fokker-Planck equation whose stationary solution describes the statistics of a laser mode around threshold showing full higher order coherence. As we shall show below, this is not the case with the solution of the complete equation (2.9a).

We solve (2.9a) in the stationary limit. As we know from the solution of the master equation, the non-diagonal part of $\rho$ vanishes, implying that $W(\alpha, \alpha^*)$ doesn't depends on the phase. Therefore we rewrite (2.9a) it in the form

$$\frac{1}{|\alpha|}\frac{d}{d|\alpha|}|\alpha|\left\{(\Gamma_0+\gamma_0-\Delta_0)|\alpha|+2(\Gamma_1-\Delta_1)|\alpha|^3+\right.$$
$$\left.\frac{1}{2}\left[\Gamma_0+\gamma_0+\Delta_0+4(\Gamma_1+\Delta_1)|\alpha|^2\right]\frac{d}{d|\alpha|}\right\}W(|\alpha|)=0 \qquad (2.10)$$

The solution of (2.10) is found by simple analytical tools and reads

$$W(\alpha,\alpha^*)=N\left(\zeta+|\alpha|^2\right)^\delta \exp\left[-\eta|\alpha|^2\right] \qquad (2.11)$$

In (2.11) we have introduced the quantities

$$\eta=\frac{(\Gamma_1-\Delta_1)}{(\Gamma_1+\Delta_1)} \qquad (2.12a)$$

$$\zeta=\frac{(\Gamma_0+\gamma_0+\Delta_0)}{4(\Gamma_1+\Delta_1)} \qquad (2.12b)$$

$$\xi=\frac{(\Gamma_0+\gamma_0+\Delta_0)(\Gamma_1-\Delta_1)}{4(\Gamma_1+\Delta_1)^2}=\eta\zeta \qquad (2.12c)$$

$$\delta=-\left(\frac{\Gamma_0+\gamma_0-\Delta_0}{2(\Gamma_1+\Delta_1)}-\xi\right) \qquad (2.12d)$$

Notice that the quantity $\delta$ defined in (2.12b) depends on $\Gamma_0+\gamma_0-\Delta_0$, which represents the gain and changes its sign as a function of the pump intensity. The normalization constant N is given by

$$N^{-1}=\frac{\exp[\xi]\,\Gamma(\delta+1,\xi)}{\eta^{\delta+1}}, \qquad (2.13)$$

where $\Gamma(\delta+1,\xi)$ is incomplete gamma function [27]. Therefore, the explicit expression for the approximate Wigner function reads

$$W(\alpha,\alpha^*) = \frac{\eta^{\delta+1}\left(\zeta+|\alpha|^2\right)^{\delta}\exp\left[-\eta\left(|\alpha|^2+\zeta\right)\right]}{\Gamma(\delta+1,\xi)} \quad , \tag{2.14}$$

which corresponds to a shifted Poisson distribution. We briefly discuss the solution (2.14) in function of the parameter $\delta$. Above threshold the gain is negative and thus $\delta > 0$. In this case the Wigner function has a maximum for $|\alpha|^2 \geq 0$. In Fig 2.1 we present the probability distributions for two values of the pump just above the threshold.

**Fig. 1. Approximate Wigner probability distribution $W(\alpha,\alpha^*)$ as a function of $|\alpha|$ for different normalized pump intensities close to the threshold and with a detuning $D = 7\ meV$. Solid line $F/F_s = 1$. Broken line $F/F_s = 1.5$. Material parameters appropriate to GaAs are used.**

The position of the maximum is given by $|\alpha|^2_{MAX} = (\delta-\zeta)/\eta$. At threshold the maximum is located at $|\alpha|^2_{MAX} = 0$. Above threshold the maximum shifts to values larger than zero indicating that the degree of coherence is growing. On the other hand, the width of the probability distribution increases as a function of the pump according to the relation $width = \eta^{-1}\sqrt{\delta/|\alpha|^2_{MAX}}$ indicating that noise effects become more relevant. These characteristics determine the behavior of the correlations as we shall see in Section III. Below threshold, the maximum disappears and the solution decreases as a function of $|\alpha|^2$. For $\delta = 0$, the solution becomes purely exponential and its width $\eta$ depends on the parametric scattering rates only. In this case the field statistics is Gaussian. Its width is determined by the fluctuations originating in the parametric scattering from the polariton ground state. The particular value of $\delta = 0$ corresponds to a situation where the ratios $\frac{\Gamma_0 + \gamma_0 - \Delta_0}{\Gamma_0 + \gamma_0 + \Delta_0}$ and $\frac{\Gamma_1 - \Delta_1}{\Gamma_1 + \Delta_1}$ of the drift and diffusion rates for non-parametric and parametric scattering respectively compensate. Finally, when $\delta < 0$, the width of the Wigner function dramatically diminishes as a function of $|\alpha|^2$.

## III. THE CORRELATION FUNCTIONS

In quantum optics the statistical properties of a radiation field are investigated experimentally by measuring either the photon number probability distribution by a photon-counting technique or its moments and the time-dependent higher order correlations in a Hanbury-Brown and Twiss experiment. In the case of polaritons, the quantity that is accessible in experiments is the photon component of the polaritons to which the quantum optical measurement techniques apply. We want to extract from the results of Section II the theoretical expression for the quantities that are accessible in the experiments. In particular, we are interested in the normalized correlations $g^{(n)}(0) = \langle P_0^{+n} P_0^n \rangle / \langle P_0^+ P_0 \rangle^n$ in the stationary regime. Using the results of Section II, the moments of the Wigner function are defined as

$$\langle P_0^{+n} P_0^n \rangle_W = \frac{1}{\pi} \int d^2\alpha |\alpha|^{2n} W(\alpha, \alpha^*) \qquad (3.1)$$

However, these quantities don't correspond to those measured in a quantum optical experiment. As it is well known, the measured moments are expressed in terms of the normal ordered moments of the polariton number distribution or of Glauber's P-function. Therefore, we have to relate the moments defined in (3.1) through the Wigner function to the experimentally relevant ones. The relation between these two different expressions for the moments is well known [22, 23] and reads

$$\langle P_0^{+n} P_0^n \rangle = \frac{1}{\pi} n! \left(-\frac{1}{2}\right)^n \int L_n^{(0)}\left(2|\alpha|^2\right) W(\alpha, \alpha^*) \, d^2\alpha \tag{3.2}$$

where the functions, $L_n^{(0)}\left(2|\alpha|^2\right)$ are the Laguerre polynomials. Using the explicit expression for the Laguerre polynomials and (3.1) we obtain from (3.2a)

$$\langle P_0^{+n} P_0^n \rangle = \sum_{m=0}^{n} (-2)^{m-n} n! \binom{n}{n-m} \frac{\langle P_0^{+m} P_0^m \rangle_W}{m!} \tag{3.3a}$$

Inserting the explicit expression of the Wigner moments obtained from (3.1) and (2.14) into (3.3) we obtain the relation

$$\langle P_0^{+m} P_0^m \rangle_W = \sum_{k=0}^{m} (-1)^k \binom{m}{k} \eta^{-k} \xi^{m-k} \frac{\Gamma(\delta+k+1,\xi)}{\Gamma(\delta+1,\xi)} \tag{3.3b}$$

The calculation of the stationary correlations $g^{(n)}(0)$ follows from (3.2) and (3.3). First of all, we test the reliability of the approximate solution (2.11) by comparing the results for the normalized second order correlation $g^{(2)}(0)$ in the stationary state calculated both from the master equation and from the approximate Wigner function.

Since from the experimental viewpoint, the values of $g^{(2)}(0)$ below threshold are not accessible due to the smallness of the signal, we report in the following examples only values of $g^{(2)}(0)$ calculated from (2.11) with $\delta > 0$. The behavior of the exact versus the approximated correlations for $\delta < 0$ sheds some light on the limits of validity of the approximation presented in Section II; this point will be discussed in detail in Section V. As a first example, we calculate the correlation $g^{(2)}(0)$ in a CdTe microcavity with the material parameters $\hbar\omega_{q=0} = E_{exc}(0) - \hbar\Omega_R$, with $E_{exc}(0) = 1680\ meV$, $2\hbar\Omega_R = 7\ meV$, $\varepsilon = 7.4$, the total exciton mass $M = 0.296$, the exciton binding energy $E_b = 25 meV$, the exciton radius $\lambda_X = 47 Å$ the quantization area $A = 10^{-5}\ cm^2$ and the detuning D = 2meV. The results are presented in Fig. 2.

**Fig. 2. Stationary normalized second order correlation function $g^{(2)}(0)$ as a function of the normalized pump intensity and a detuning *D=2 meV*. Solid line: result obtained with the approximate Wigner function (2.14). Broken line: results obtained from the master equation (2.5). Dots: experimental points [10]. The vertical line indicates the position of the threshold. Material parameters appropriate to CdTe are used.**

where the experimental points from [10] have been introduced and shall be discussed later. From Fig. 2, we see that the correlation calculated with the approximate solution reproduces the one calculated with the master equation (2.5) with a high accuracy for most values of the external pump. The differences found for large values of the pump are due to the choice of the number of steps in calculating the numerical solution of (2.5). In order to obtain the same accuracy that is found for smaller values of the pump, the number of steps in the numerical calculation should be substantially raised, stretching the already long calculation time. On the contrary, calculating with the approximate solution requires only a few seconds.

Since the most recent experiments [11] have been performed in GaAs microcavities, we present in Figs. 3.2 $g^{(2)}(0)$ for different detunings (2 and 7 $meV$) showing some peculiarities of the correlation function. The material parameters for GaAs microcavities are $E_{exc}(0) = 1509\ meV$, $\varepsilon = 12.4$, the total exciton mass $M = 0.517 m_e$, the exciton binding energy $E_b = 10 meV$, the exciton radius $\lambda_X = 50 Å$ the quantization area $A = 10^{-5} cm^2$ with detuning of 2 $meV$ and of 7 $meV$.

**Fig. 3. Stationary normalized second order correlation function $g^{(2)}(0)$ as a function of the normalized pump intensity evaluated with the approximate Wigner function (2.14) for two different detunings *D*. Broken line *D*= 2 *meV*. Solid line *D*= 7 *meV*. The insert indicates that a reduced minimum is also present for *D*= 2 *meV*. The vertical line indicates the position of the threshold. Material parameters appropriate to GaAs are used.**

We notice that, when crossing the threshold $g^{(2)}(0)$ drops from the incoherent value $g^{(2)}(0) = 2$ to a value close to $g^{(2)}(0) = 1$ that is characteristic of a full coherent field, as already pointed out in [14, 16]. However, for larger values of the pump the correlation grows again. This behavior is due to the presence of the parametric noise i.e. the parametric diffusion coefficient (2.9c) as we shall show later. The minimum in $g^{(2)}(0)$ depends on the detuning as well as on the material parameters.

The minimum in Fig. 3. in the case of detuning $D = 2\ meV$ has a very small depth and $g^{(2)}(0)$ shows a slow increasing as a function of the external pump. On the contrary, for detuning $D = 7 meV$ the minimum is analogous to the one presented in Fig. 2 for CdTe. This indicates that the smaller detuning implies a larger influence of the parametric scattering on the statistics of the condensate. The same consideration hold in the case of a CdTe microcavity when the detuning is changed from $D = 0\ meV$ to $D = 2 meV$. We also notice, that the difference in the minima of $g^{(2)}(0)$ that is observed between CdTe and GaAs microcavities with detuning $D = 2 meV$ arises form the differences in the material parameters.

We get a better understanding of the insurgence of the minimum in $g^{(2)}(0)$ by relating it to the behavior of the corresponding probability distribution. The dependence of this minimum on detuning is also understood in terms of the behavior of the probability distributions. In Fig. 4. we compare the probability distributions of the polaritons in

**Fig. 4. Approximate Wigner probability distribution $W(\alpha,\alpha^*)$ as a function of $|\alpha|$ for different normalized pump intensities and detunings. Grey solid line $F/F_s = 1$, $D = 7\ meV$. Black solid line $F/F_s = 2$, $D = 7\ meV$. Grey broken line $F/F_s = 1$, $D = 2\ meV$. Black broken line $F/F_s = 1.5$, $D = 2\ meV$. Material parameters appropriate to GaAs are used.**

GaAs corresponding to the detunings $D = 2meV$ and $D = 7meV$ respectively, just above threshold. We notice that the shift of the maximum of the probability distribution is larger in the $D = 7meV$ case than in the $D = 2meV$ case. On the contrary, its width is larger in the $D = 7meV$ case than in the $D = 2meV$ case. As pointed out in Section II, a larger width of $W(\alpha,\alpha^*)$ combined with a smaller shift of its maximum implies a lower degree of coherence. This fact explains the difference in the depth of the minima in Fig. 3.

When comparing with the experiments, in the case of CdTe, as shown in Fig. 2 the theoretical results are in a qualitative agreement with the experimental ones [10], a quantitative agreement being found only for large values of the pump and near the threshold. The difference in the growth of $g^{(2)}(0)$ is related to the magnitude of the parametric diffusion rate, which is apparently overestimated in our model. The agreement of our numerical results with the measurements performed in GaAs with detuning of 2 $meV$ [11] is fair, as sown in Fig. 5. for values of the pump not too closed to the threshold and for both the second and the third order correlation.

**Fig. 5. Stationary normalized second order correlation function $g^{(2)}(0)$ as a function of the normalized pump intensity. Dots: experimental results[11]. Solid line: result obtained with the approximate Wigner function (2.14) and with material parameters appropriate to GaAs. The vertical line indicates the position of the threshold.**

Due to the low intensity of the signal just above threshold, the first few points may not be significant. Furthermore, as pointed out in [11], the discrepancy between the theoretical and experimental values near the threshold may be related to the measurement technique. Notice that no fit procedure in used in Figs. 3.1 and 3.4 when comparing theory and experiment.

We conclude this Section by briefly discussing the behavior of the field amplitude in the context of the Fokker-Planck description. As we already know, the diagonal and off-diagonal matrix elements of the polariton density operator in (2.5) evolve separately. Therefore, the off-diagonal matrix elements of the density operator and hence the expectation value of the polariton amplitude vanishes for any time, if it vanishes for t=0. In the framework of the Fokker-Planck description this result indicates that the polariton field in the stationary regime doesn't depend on the phase. However, its amplitude may have a stationary value different from zero. This is indeed the case as it is shown by using the stationary solution (2.14) and the definition

$$\langle |P_0| \rangle_W = \int_0^\infty d|\alpha|^2 |\alpha| W(\alpha, \alpha^*)$$

The result is presented in Fig. 6 and compared with the results from the master equation (2.5). We notice that once more the approximate solution leads to an excellent result when compared with the results from the master equation.

**Fig. 6. Modulus of the amplitude as a function of the normalized pump intensity. Solid line: result obtained with the approximate Wigner function (2.14). Broken line: result obtained from the master equation (2.5). The vertical line indicates the**

**position of the threshold. The difference between the two approaches is relevant only for high intensities as visualized in the insert. Material parameters appropriate to CdTe are used.**

## IV. THE LANGEVIN EQUATION

In this Section we shall consider some characteristics of the time evolution of the ground-state polaritons starting from the time-dependent Fokker-Planck equation (2.9). In general, the coefficients in (2.9) are time dependent quantities. Since we are interested in the solution of (2.9) close to the stationary regime, we assume that the coefficients in (2.9) take their stationary values. In spite of this simplification, a numerical solution of (2.9) is cumbersome and we shall not discuss it here. In order to describe the time dependent behavior of the condensate polaritons close to the stationary state, we have chosen to work in a different and simpler calculation scheme. It consists in going over from a Fokker-Planck to a Langevin description of the dynamics. Since the two approaches are equivalent, we don't loose any information in this new description. The Langevin equations for the polariton field amplitude are obtained from (2.9) using standard methods (see e.g. Ref. [26]) and read

$$\hbar \frac{d}{dt}\alpha = -i\hbar\omega_0\alpha - \left[\Gamma_0 + \gamma_0 - \Delta_0 - 2(\Gamma_1 + \Delta_1)\right]\alpha - \left[2(\Gamma_1 - \Delta_1) + 2iW_{0,0,0}\right]|\alpha|^2 \alpha + F(t) \quad (4.1)$$

$$\hbar \frac{d}{dt}\alpha^* = i\hbar\omega_0\alpha^* - \left[\Gamma_0 + \gamma_0 - \Delta_0 - 2(\Gamma_1 + \Delta_1)\right]\alpha^* - \left[2(\Gamma_1 - \Delta_1) - 2iW_{0,0,0}\right]|\alpha|^2 \alpha^* + F^*(t) \quad (4.2)$$

In (4.1) and (4.2), $F(t)$ denotes for the stochastic Langevin force defined through the correlations

$$\langle F(t) \rangle = 0 \tag{4.3a}$$

$$\langle F^*(t')F(t) \rangle = \left( 2\left[ \Gamma_0 + \gamma_0 + \Delta_0 \right] + 8\left( \Gamma_1 + \Delta_1 \right)|\alpha|^2 \right) \delta(t-t') \tag{4.3b}$$

$$\langle F(t')F(t) \rangle = \langle F^*(t')F^*(t) \rangle = 0 \tag{4.3c}$$

The higher order correlations of the Langevin force factorize consistently with (4.3). Calculations of the relaxation times are more conveniently performed rewriting the Langevin equations using the absolute value $|\alpha|$ and the phase φ as variables. Indeed, the expectation value of the polariton amplitude α vanishes in the stationary regime while the expectation value of its absolute value is different from zero in the same regime as shown in Section III. Therefore, we expect the dynamics of the phase to be responsible for the vanishing of the polariton amplitude in the stationary regime. By separating modulus and phase in (4.1) and (4.2), we obtain

$$\hbar \frac{d}{dt}|\alpha| = -\left[ \Gamma_0 + \gamma_0 - \Delta_0 - 2(\Gamma_1 + \Delta_1) \right]|\alpha| -$$
$$\left[ 2(\Gamma_1 - \Delta_1) \right]|\alpha|^3 + \Phi(t) \tag{4.4a} \text{ I}$$

$$\hbar \frac{d}{dt}\varphi = -\hbar\omega_0 - 2W_{0,0,0}|\alpha|^2 + \Psi(t) \tag{4.4b}$$

n (4.4), $\Phi(t)$ and $\Psi(t)$ denote the stochastic Langevin forces associated with amplitude and phase that are related to the stochastic forces $F(t)$ and whose second order correlations are

$$\langle \Phi(t')\Phi(t)\rangle =$$
$$\frac{1}{4}\langle \left(F(t')\exp(-i\varphi(t'))+F^*(t')\exp(i\varphi(t'))\right)\left(F(t)\exp(-i\varphi(t))+F^*(t)\exp(i\varphi(t))\right)\rangle$$
$$\langle \Psi(t')\Psi(t)\rangle =$$
$$\frac{1}{4}\langle \left(F(t')\exp(-i\varphi(t'))+F^*(t')\exp(i\varphi(t'))\right)\left(F(t)\exp(-i\varphi(t))-F^*(t)\exp(i\varphi(t))\right)\rangle$$

Since we are interested in the time-dependent behavior of the polaritons close to the stationary regime, we linearize (4.4) around the stationary expectation value $\langle|\alpha|\rangle_S$ of the amplitude and obtain

$$\hbar\frac{d}{dt}|\alpha|_{lin} = -\left[\Gamma_0+\gamma_0-\Delta_0-2(\Gamma_1+\Delta_1)\right]|\alpha|_{lin} - $$
$$\left[2(\Gamma_1-\Delta_1)\right]\langle|\alpha|\rangle_S^2 |\alpha|_{lin} + \Phi(t) \qquad (4.5a)$$

$$\hbar\frac{d}{dt}\varphi = -\hbar\omega_0 - 2W_{0,0,0}\left(|\alpha|_{lin}\langle|\alpha|\rangle_S + \langle|\alpha|\rangle_S^2\right) + \Psi(t) \qquad (4.5b)$$

where
$$|\alpha|_{lin} = |\alpha| - \langle|\alpha|\rangle_S$$

Integrating (4.5a) in time and taking its expectation value we obtain

$$\langle|\alpha|\rangle_{lin}(t) = \langle|\alpha|\rangle_{lin}(0)\exp(-\Gamma t/\hbar) \qquad (4.6a)$$
with
$$\Gamma = \left[\Gamma_0+\gamma_0-\Delta_0-2(\Gamma_1+\Delta_1)\right]+\left[2(\Gamma_1-\Delta_1)\right]\langle|\alpha|\rangle_S^2 \qquad (4.6b)$$

Since by definition $\langle|\alpha|\rangle_{lin}(0)=0$, the contribution of $\langle|\alpha|\rangle_{lin}(t)$ to the time evolution close to the stationary state doesn't play any role. On the contrary, the time evolution of the phase determines the decay characteristics of the polariton amplitude, as we shall see in the following. As a consequence of the above considerations equation (4.5b) reads

$$\hbar\frac{d}{dt}\varphi = -\hbar\omega_0 - 2W_{0,0,0}\langle|\alpha|\rangle_S^2 + \Psi(t) \qquad (4.7)$$

We integrate (4.7) in time and obtain for the field amplitude the expression

$$\alpha(t) = \langle |\alpha| \rangle_s \exp(i\varphi(t)) =$$

$$= \langle |\alpha| \rangle_s \exp\left(-i\omega_0 t - 2iW_{0,0,0}\langle |\alpha| \rangle_S^2 t/\hbar + i\int_0^t \Psi(t')dt'/\hbar\right) \qquad (4.8)$$

From (4.8) we obtain the first order time dependent correlation of the polariton field i.e.

$$\langle \alpha^*(t)\alpha(0) \rangle = G^{(1)}(t) = \langle |\alpha|^2 \rangle_S \exp\left[-i\left(\omega_0 + 2W_{0,0,0}\langle |\alpha|^2 \rangle_S/\hbar\right)t - \gamma_c t/\hbar\right] \qquad (4.9a)$$

with

$$\gamma_c = \left(\frac{\left([\Gamma_0 + \gamma_0 + \Delta_0] + 4(\Gamma_1 + \Delta_1)\langle |\alpha|^2 \rangle_S\right)}{\langle |\alpha|^2 \rangle_S}\right) \qquad (4.9b)$$

where $\gamma_c$ is the linewidth of the polariton in the ground state.

Taking the time Fourier transform of (4.9) we obtain the spectrum of the fluctuations of the polariton amplitude near the stationary situation. It reads

$$S^{(1)}(\omega) = \frac{G^{(1)}(0)}{i\left(\hbar\omega - \hbar\omega_0 + 2W_{0,0,0}\langle |\alpha| \rangle_S^2\right) + \gamma_c} \qquad (4.9c)$$

First of all, we notice that the linewidth is determined by the diffusion coefficient. This fact is not surprising because, as we shall show below, the phase undergoes a diffusion process in the time domain considered here. From (4.9) it follows that in the considered regime the dynamics of $|\alpha|$ and of $\varphi$ evolve separately. Furthermore, close to the stationary solution only the time dependence of the phase is relevant. Because of this behavior, we may factorize the Wigner function in this regime as $W(\alpha, \alpha^*, t) = W_{stat}(|\alpha|)W_\varphi(\varphi, t)$ and rewrite (2.9) by separating the polariton amplitude in absolute value and phase. After inserting the above Ansatz into (2.9) and integrating over the absolute value of the amplitude, it is found that $W_\varphi(\varphi,t)$ obeys the following diffusion equation,

$$\hbar \frac{\partial W_\varphi(\varphi,t)}{\partial t} = -\left(2\hbar\omega_0 + 2W_{0,0,0}\langle|\alpha|^2\rangle_s\right)\frac{\partial}{\partial \varphi}W_\varphi(\varphi,t) + \qquad (4.10)$$

$$\frac{\partial}{\partial \varphi^2}\left(\left\langle\frac{1}{|\alpha|^2}\right\rangle_s (\Gamma_0 + \gamma_0 + \Delta_0) + 4(\Gamma_1 + \Delta_1)\right)W_\varphi(\varphi,t)$$

The explicit expression (4.9) of the linewidth shows that it consists in the interplay of two terms. The first term is inversely proportional to the polariton intensity and thus becomes smaller when the polariton population increases. The second term is determined by the sum of the dissipation and injection rates due to the parametric scattering and its value grows with the pump. For large values of the external pump, when the polariton population is large, this term dominates. Therefore, we expect that the linewidth strongly decreases as a function of the external pump when approaching the threshold and starts growing above threshold due to the contribution of the parametric scattering. Indeed, this behavior of the linewidth is obtained as shown in Fig. 7 using CdTe material parameters.

**Fig. 7. Solid line: plot of the linewidth $\gamma_c$ (Eq. 4.9b) as a function of the normalized pump intensity. Broken line: plot, as a function of the normalized pump intensity, of the parametric diffusion $(\Gamma_1 + \Delta_1)$ (the sum of injection and dissipation rates),**

**which dominates $\gamma_c$ for high pump intensities. The vertical line indicates the position of the threshold. Material parameters appropriate to CdTe are used.**

The values for the linewidth obtained from (4.9b) are comparable in magnitude with the one obtained experimentally both for a CdTe [10] and for a GaAs microcavity [11]. However, the dependence on the pump intensity shown in Fig. 7 only qualitatively reproduces the experimental curves. This discrepancy is related to the experimental limitations as well as to the simplicity of the theoretical description of the incoherent processes. In Fig. 7 we also show explicitly that above threshold the growth of the linewidth with the pump is mainly due to the presence of the parametric noise as already discussed above.

We obtain the second order time-dependent correlation close to the stationary state $G^{(2)}(t) = \langle \alpha^*(0)\alpha^*(t)\alpha(t)\alpha(0) \rangle_S$ through an analogous calculation not involving the phase. From (4.4) it follows

$$\hbar \frac{d}{dt} G^{(2)}(t) = -2\left[\Gamma_0 + \gamma_0 - \Delta_0 - 2(\Gamma_1 + \Delta_1)\right] G^{(2)}(t) -$$
$$4\left[2(\Gamma_1 - \Delta_1)\right]\langle|\alpha|^2\rangle_S G^{(2)}(t) +$$
$$2\left(\langle|\alpha(0)||\alpha_L(t)|\Phi(t)|\alpha(0)|\rangle\right) \quad (4.11)$$

In order to integrate this linear equation we need to specify the expectation value of the product between the stochastic force and the polariton amplitudes that is given by

$$\langle |\alpha(0)||\alpha_{lin}(t)||\alpha(0)|\Phi(t)\rangle =$$

$$\left\langle |\alpha(0)|\int_0^t dt'\exp\left(-i(\Omega-i\Gamma/\hbar)(t-t')\right)\Phi(t')\Phi(t)|\alpha(0)|/\hbar\right\rangle =$$

$$\left(2[\Gamma_0+\gamma_0+\Delta_0]\langle|\alpha|^2\rangle_S + 8(\Gamma_1+\Delta_1)G^{(2)}(t)\right) \tag{4.12}$$

where $\Gamma$ is defined in (4.6b) and $\Omega = \omega_0 + 4W_{0,0,0}\langle|\alpha|^2\rangle_S/\hbar$

Inserting (4.12) into (4.11) we obtain

$$\hbar\frac{d}{dt}G^{(2)}(t) = -2\left[\Gamma_0+\gamma_0-\Delta_0-2(\Gamma_1+\Delta_1)\right]G^{(2)}(t) -$$
$$4\left[(\Gamma_1-\Delta_1)\right]\langle|\alpha|^2\rangle_S G^{(2)}(t) + 8(\Gamma_1+\Delta_1)G^{(2)}(t) +$$
$$2(\Gamma_0+\gamma_0+\Delta_0)\langle|\alpha|^2\rangle_S \tag{4.13}$$

We notice that the width of the second order time correlation depends as well on the non-parametric processes leading to the gain as on the correlations of the Langevin force that are related to the parametric processes.. The spectrum of the second order correlation is

$$S^{(2)}(\omega) = \frac{2(\Gamma_0+\gamma_0+\Delta_0)\langle|\alpha|^2\rangle_S + G^{(2)}(0)}{i\hbar\omega + \left[2\Gamma - 8(\Gamma_1+\Delta_1)\right]} \tag{4.14}$$

The linewidth appearing in (4.14) is related to the inverse of the relaxation time of the emitted intensity. In the following we shall call it relaxation rate. We notice that the dependence of the relaxation rate on the scattering rates and on the stationary population differs from the linewidth of the phase correlations. In the case of the second order correlations, the leading contribution to the relaxation rate originates in the gain rate $2\left[\Gamma_0+\gamma_0-\Delta_0-2(\Gamma_1+\Delta_1)\right]$ and in the saturation rate $4\left[(\Gamma_1-\Delta_1)\right]\langle|\alpha|^2\rangle_S$. Below threshold the gain is positive and becomes zero at threshold thus implying that the relaxation rate strongly decreases below threshold. Notice, that it doesn't become zero because of the presence of the saturation term and of the contribution originating in the correlation of the Langevin forces. Above threshold the gain becomes negative but the saturation term grows with the polariton intensity and guarantees the growing of the decay rate as a function of the pump. This behavior is similar to the one that is found in a conventional laser. The contribution of the noise competes with the gain rate in reducing the growth of the relaxation rate above threshold with respect to the laser case. The behavior discussed above is presented in Fig. 8 for CdTe material parameters.

**Fig. 8. Solid line: plot of the relaxation rate of the polaritons (see Eq. 4.14) as a function of the normalized pump intensity. Broken line: plot of the relaxation rate where of the parametric diffusion $(\Gamma_1 + \Delta_1)$ (the sum of injection and dissipation rates) is neglected. The insert indicates the decreasing of the relaxation rate below threshold. The vertical line indicates the position of the threshold. Material parameters appropriate to CdTe are used.**

Here too the values for the relaxation rate obtained from (4.13) are of the same order of magnitude as the one measured in the experiments but only qualitatively reproduce the experimental results. In Fig. 8, we also compare the relaxation rate of the polaritons with the one obtained by neglecting the effects of the polariton noise. With growing pump the effects of the parametric diffusion on the relaxation rate grow.

## V. THE EXACT WIGNER FUNCTION

In this Section we present an analytical solution of the complete equation (2.8) derived in Section II in the stationary regime. We may ask what is the relevance of the solution of (2.8) in the context of the present approach. Indeed, as we shall see below, the calculation of the correlations using the exact analytical solution is in general a numerically difficult task and doesn't bring much advantage with respect to solving the master equation directly. Furthermore, from the viewpoint of the applications, the approximate solution (2.11) leads to very good results for $\delta > 0$ and there is no need for exact but more cumbersome calculations. Therefore, the exact analytic solution of (2.8) seems to be only of academic relevance. However, the exact analytical solution of (2.8) allows us to obtain a quantitative understanding of the range of validity of the approximate solution discussed in the Section II. This point is of general interest because, as already mentioned, neglecting the third order derivatives in (2.8) corresponds to what in the theory of Bose-Einstein condensation is called the truncated Wigner method [19]. Therefore, a quantitative discussion of the limits of validity of the approximation performed in Section II may also shed some light on its limits of validity in other contexts.

In order to simplify the following steps we rewrite here the complete equation (2.8) for the Wigner function by separating absolute value and phase of the polariton field obtaining

$$\hbar \frac{\partial W(\alpha,\alpha^*)}{\partial t} = \frac{1}{|\alpha|}\frac{\partial}{\partial |\alpha|}|\alpha|\left\{2[\Gamma_0+\gamma_0-\Delta_0]|\alpha|+4(\Gamma_1-\Delta_1)|\alpha|^3\right\}W(\alpha,\alpha^*)+$$

$$\frac{1}{|\alpha|}\frac{\partial}{\partial |\alpha|}|\alpha|\left\{\left[\Gamma_0+\gamma_0+\Delta_0+4(\Gamma_1+\Delta_1)|\alpha|^2\right]\frac{\partial}{\partial |\alpha|}\right\}W(\alpha,\alpha^*)+$$

$$\frac{1}{|\alpha|}\frac{\partial}{\partial |\alpha|}|\alpha|\left\{(\Gamma_1-\Delta_1)\left(|\alpha|\frac{\partial^2}{\partial |\alpha|^2}+\frac{\partial}{\partial |\alpha|}\right)\right\}W(\alpha,\alpha^*)+$$

$$L\bigl[|\alpha|,\varphi\bigr]W(\alpha,\alpha^*) \qquad (5.1)$$

where the phase-dependent part of the differential operator in (5.1) is

$$L\bigl[|\alpha|,\varphi\bigr]W(\alpha,\alpha^*) = \left[(\Gamma_1-\Delta_1)\frac{1}{|\alpha|^3}\frac{\partial^2}{\partial \varphi^2}\left(\frac{\partial}{\partial |\alpha|}|\alpha|\right)\right]W(\alpha,\alpha^*)+$$

$$\left[2W_{0,0,0}\frac{\partial}{\partial \varphi}\left(\frac{1}{|\alpha|}\frac{\partial}{\partial |\alpha|}|\alpha|\frac{\partial}{\partial |\alpha|}+\frac{1}{|\alpha|^2}\frac{\partial^2}{\partial \varphi^2}+|\alpha|^2\frac{\partial}{\partial \varphi}\right)\right]W(\alpha,\alpha^*)+$$

$$\left[\Gamma_0+\gamma_0+\Delta_0-2(\Gamma_1-\Delta_1)+4(\Gamma_1+\Delta_1)|\alpha|^2\right]\frac{1}{|\alpha|^2}\frac{\partial^2}{\partial \varphi^2}W(\alpha,\alpha^*)+$$

$$2\hbar\omega_0\frac{\partial}{\partial \varphi}W(\alpha,\alpha^*)$$

The stationary solution is assumed to be independent of φ and is determined by

$$\left[(\Gamma_0+\gamma_0-\Delta_0)|\alpha|+2(\Gamma_1-\Delta_1)|\alpha|^3\right]W\bigl(|\alpha|^2\bigr)+$$

$$\frac{1}{2}\left[\Gamma_0+\gamma_0+\Delta_0+(\Gamma_1-\Delta_1)+4(\Gamma_1+\Delta_1)|\alpha|^2\right]\frac{dW\bigl(|\alpha|^2\bigr)}{d|\alpha|}+$$

$$|\alpha|\frac{(\Gamma_1-\Delta_1)}{2}\frac{d^2W\bigl(|\alpha|^{2*}\bigr)}{d|\alpha|^2}=0 \qquad (5.2)$$

Notice that the equation (2.9b), from which to the approximate solution is calculated, follows from (5.2) by neglecting the second order derivative as well as the term $(\Gamma_1-\Delta_1)$ in the coefficient of the first order derivative. Both these terms originate in the third order derivative in (2.8). The equation (5.2) is a second order differential equation that is easily put into a standard form. To this end, we change the variable $|\alpha|^2$ into x, and introduce the definitions (2.12) obtaining

$$x\frac{d^2W(x)}{dx^2}+\left(\frac{4\zeta+\eta}{\eta}+\frac{4}{\eta}x\right)\frac{dW(x)}{dx}+\left(\frac{4\zeta-4\delta}{\eta}+4x\right)W(x)=0 \qquad (5.3)$$

We then introduce the Ansatz

$$W(x)=\exp(-\mu x)Z(x) \qquad (5.4)$$

into (5.3), leading to a differential equation for $Z(x)$. The constant $\mu$ is determined such that the coefficient of $Z(x)$ in the resulting differential equation is independent of $x$. We obtain

$\mu_{1,2}=2\left(\frac{1}{\eta}\pm\sqrt{\beta}\right)$, where $\beta=\left(\frac{1}{\eta^2}-1\right)$. After choosing the positive root for $\mu$ and

introducing the new variable $z=4\sqrt{\beta}x$ we obtain

$$z\frac{d^2Z(z)}{dz^2}+(b-z)\frac{dZ(z)}{dz}-aZ(z)=0 \qquad (5.5a)$$

with

$$a=\frac{\delta-\xi}{\eta\sqrt{\beta}}+\frac{4\zeta+\eta}{2\eta}+\frac{4\zeta+\eta}{2\eta^2\sqrt{\beta}}, \qquad (5.5b)$$

$$b=\frac{4\zeta}{\eta}+1. \qquad (5.5c)$$

Equation (5.5a) is the well-known Kummer equation, whose two linearly independent solutions are the confluent hypergeometric functions $M(a,b,z)$ and $U(a,b,z)$ [27]. Since the solution must converge and be continuous when $z \to 0$, we obtain for the exact Wigner function, which from here on will be denoted by $W^{exact}(z)$ to avoid confusion with the approximate one, the expression

$$W^{exact}(z) = N \exp\left(-\frac{z}{2\eta\sqrt{\beta}}\right) \exp(-z/2) M(a,b,z) \qquad (5.6)$$

The solution (5.6) exists, provided that the condition $\beta > 0$ i.e. $(\Gamma_1 + \Delta_1) > (\Gamma_1 - \Delta_1)$ is satisfied, which is always the case.. The normalization in (5.6) is defined as

$$N^{-1} = \int_0^\infty \exp\left(-\frac{z}{2\eta\sqrt{\beta}}\right) \exp(-z/2) \frac{M(a,b,z)}{4\sqrt{\beta}} dz = \Gamma(1) \frac{{}_2F_1(1;a;b;1/A)}{4\sqrt{\beta}A} \qquad (5.7a)$$

$$A = \frac{1+\eta\sqrt{\beta}}{2\eta\sqrt{\beta}} \qquad (5.7b)$$

provided that $A > 1$. Here ${}_2F_1(1,a,b,1/A)$ is the hypergeometric function.

In order to have some insight in the behavior of the approximate Wigner function (2.11) and of the exact one (5.6), we consider these quantities as functions of the parameters $\delta$ and $(a,b)$ respectively. The approximate Wigner function for $\delta = 0$ takes the form $W(x; \delta = 0) = \eta \exp[\eta x]$, moreover defining $G = \Gamma_0 + \gamma_0 - \Delta_0$ and $Q = \Gamma_0 + \gamma_0 + \Delta_0$, $\delta = 0$ implies $G = \frac{Q}{2}\left(\frac{\Gamma_1 - \Delta_1}{\Gamma_1 + \Delta_1}\right)$. The exact Wigner function for $a=b$ becomes

$$W^{exact}(x; a=b) = \frac{2-2\eta\sqrt{\beta}}{\eta} \exp[-\frac{2-2\eta\sqrt{\beta}}{\eta} x], \text{ which for } \eta = 1 \text{ reduces to}$$

$W_{approx}(x; \delta = 0)$. In this case $a=b$ implies $G = \dfrac{Q}{2}\left(\dfrac{\Gamma_1 - \Delta_1}{\Gamma_1 + \Delta_1}\right) + \dfrac{1}{2}\left(\dfrac{(\Gamma_1 - \Delta_1)^2}{\Gamma_1 + \Delta_1}\right)$. The correction $\dfrac{1}{2}\left(\dfrac{(\Gamma_1 - \Delta_1)^2}{\Gamma_1 + \Delta_1}\right)$ being small for $\eta = 1$.

Furthermore, we are able to obtain some information on the behavior of the exact solution from the following asymptotic expansion of the function $M(a,b,z)$ [27] valid for $|z| \to \infty$:

$$M(a;b;z) = \dfrac{\Gamma(a)}{\Gamma(b)} z^{a-b} \exp(z) + O\left(\dfrac{1}{|z|}\right), \qquad (5.9a)$$

which leads to the expression for the function $W^{exact}(z)$

$$W^{exact}(z) = \dfrac{\Gamma(a)}{\Gamma(b)} z^{a-b} \exp\left[-z\left(\dfrac{1}{2\eta\sqrt{\beta}} + \dfrac{1}{2}\right)\right] + O\left(\dfrac{1}{|z|}\right). \qquad (5.9b)$$

Reintroducing the original variable $x = |\alpha|^2 = z/4\sqrt{\beta}$, we obtain for $(|\alpha| \gg 1, \eta \ll 1)$ the asymptotical Wigner function $W^{eact}(\alpha, \alpha^*)$: $|\alpha|^{2\delta} \exp\left[-\eta|\alpha|^2\right]$. Therefore, the asymptotic expansions for the exact and for the approximate solutions are the same. Since for $x = 0$ both the exact and the asymptotic solution have very small although different values, we conclude that for $\delta > 0$ both solutions qualitatively have the same behavior as functions of $x$ showing a maximum for $x \neq 0$ above threshold. When $\delta < 0$, both the exact and the approximate Wigner function decrease as functions of $x$. However, their behavior differs for small values of $x$. In fact the exact solution $\exp(-z/2\eta\sqrt{\beta})\exp(-z/2)M(a,b,z)$ takes the value one for $z = 0$ while the function $(z/4\sqrt{\beta}+\zeta)^\delta \exp(-\eta z/4\sqrt{\beta})$ takes the value $\zeta^\delta$ for $z=0$, which is very small. This difference is crucial when calculating the integrals leading to the moments of the probability distribution. In the asymptotic limit $x \gg 1$ both solutions have the same behavior as for $\delta > 0$.

A quantitative comparison of the solutions is not presented here. In fact, the exact solution (5.6) for the Wigner function poses some computational problems related to the numerical values of the relaxation and injection rates, which depend on the material parameters. From the results of the Boltzmann dynamics, the parametric scattering rates $\Gamma_1$, $\Delta_1$ are found to be of the same order of magnitude and thus $\eta = 1$. As a consequence, the values of the parameters $a$ and $b$ in the confluent hypergeometric function $M(a,b,z)$ as well as its variation scale in $x$ are large. In this case, the numerical evaluation of (5.6) becomes cumbersome.

The moments are calculated through the integral

$$\langle P_0^{+n} P_0^n \rangle_W^{exact} = N \int_0^\infty \exp\left(-\frac{z}{2\eta\sqrt{\beta}}\right) \frac{z^n \exp(-z/2)\, M(a,b,z)}{\left(4\sqrt{\beta}\right)^{n+1}} dz =$$
$$N\Gamma(n+1) \frac{{}_2F_1(n+1,a,b,1/\alpha)}{\left(4\sqrt{\beta}\right)^{n+1} A^{n+1}} \tag{5.8}$$

The calculation of the moments of the Wigner function $W^{exact}(z)$ through (5.8) has been performed with CdTe material parameters leading to results that coincide with the ones obtained from the numerical solution of (2.5) as expected. However, with GaAs material parameters the same calculations become cumbersome. This result shows the advantage of using the approximate solution, which leads to simple calculations for both choices of the material parameters.

We can exploit the relations between the normally ordered and approximated moments expressed in (3.3a) in order show that indeed the moments calculated from the exact or of approximate solution (2.14) show intrinsic differences. By multiplying (2.8) by $|\alpha|^2$ and integrating over $\alpha$ we obtain the equation governing the evolution of the first Wigner moment. Using (3.3a) it is then transformed into an equation for the normal ordered moments, which reads

$$\hbar \frac{d}{dt}\langle P_0^+ P_0 \rangle^{exact} = -2\left\{(\Gamma_0 + \gamma_0 - \Delta_{0,}) - 8\Delta_1\right\}\langle P_0^+ P_0 \rangle^{exact} -$$
$$4(\Gamma_1 - \Delta_1)\langle P_0^{+2} P_0^2 \rangle^{exact} + 2\Delta_{0,} + 8\Delta_1 \qquad (5.10a)$$

whereas the same equation with the approximation (2.11), leads to

$$\frac{d}{dt}\langle P_0^+ P_0 \rangle^{approx} = -2\left\{(\Gamma_0 + \gamma_0 - \Delta_{0,}) - 8\Delta_{11}\right\}\langle P_0^+ P_0 \rangle^{approx} -$$
$$4(\Gamma_1 - \Delta_1)\langle P_0^{+2} P_0^2 \rangle^{approx} + 2\Delta_{0,} + 8\Delta_1 - 2(\Gamma_1 - \Delta_1) \qquad (5.10b)$$

The equations for the first moments differ only in the last term representing the contribution of the parametric diffusion, which in our model leads to a small correction. In the same way we obtain the equations for the second moment for both the exact and the approximate case that read

$$\hbar \frac{d}{dt}\langle P_0^{+2} P_0^2 \rangle^{exact} = -4(\Gamma_0 + \Gamma_1 + \gamma_0 - 13\Delta_1 - \Delta_0)\langle P_0^{+2} P_0^2 \rangle^{exact} -$$
$$8(\Gamma_1 - \Delta_1)\langle P_0^{+3} P_0^3 \rangle^{exact} +$$
$$(64\Delta_1 + 8\Delta_0)\langle P_0^+ P_0 \rangle^{exact} + 8\Delta_1 \qquad (5.11a)$$

and

$$\frac{d}{dt}\langle P_0^{+2} P_0^2 \rangle^{approx} = -4(\Gamma_0 + \Gamma_1 + \gamma_0 - 13\Delta_1 - \Delta_0)\langle P_0^{+2} P_0^2 \rangle^{approx} -$$
$$8(\Gamma_1 - \Delta_1)\langle P_0^{+3} P_0^3 \rangle^{approx} +$$
$$(64\Delta_1 + 8\Delta_0)\langle P_0^{+} P_0 \rangle^{approx} + 8\Delta_1 -$$
$$4(\Gamma_1 - \Delta_1)\langle P_0^{+} P_0 \rangle^{approx} + 2(\Gamma_1 - \Delta_1) \tag{5.11b}$$

The equations (5.11a) and (5.11b) differ in the last two terms of (5.11*b*) that depend on the parametric scattering rates. Analogous expressions are found for the higher order moments. The relevance of these corrections is made evident by comparing the moments calculated with the approximate Wigner function (2.11) and with the exact solution that for simplicity we extract from the master equation (2.5). The behavior of the moments is different depending on the sign of the quantity $\delta$ in the approximate solution. For $\delta > 0$, the difference between the exact and the approximate moments is very small at least for the first ten moments. This result follows from the fact that the exact and the approximate solution, as pointed out above, have qualitatively the same behavior as functions of x for $\delta > 0$. Since the probability distributions are determined by their moments, we expect to find slightly larger differences between both solutions for very high order moments only. When $\delta = 0$ the exact and approximate Wigner functions coincide as shown above. When $\delta < 0$ the situation becomes more involved. With growing order of the moment, the values of the approximated moments become less reliable and from the fifth order moment on, no values of the moments for $\delta < 0$ are acceptable. From the numerical viewpoint, this fact is related to the smallness of the quantities involved in the calculations of the approximate moments. From a conceptual viewpoint, the differences between the exact and the approximated moments that grow with the order of the moment express the fact that the approximated and exact solutions correspond to different probability distributions. We conclude that in the present model the truncated Wigner method leads to very good results above threshold and in the region below threshold where $\delta \geq 0$. The truncated Wigner method becomes questionable when $\delta < 0$ i.e. below threshold and for small values of the pump intensity. This regime corresponds to the one characterized by a polariton population $\langle P_0^+ P_0 \rangle < \frac{\Gamma_1 + \Delta_1}{\Gamma_1 - \Delta_1}$. Here $\frac{\Gamma_1 + \Delta_1}{\Gamma_1 - \Delta_1}$ is the polariton population at $\delta = 0$ where the approximate and the exact Wigner functions coincide. This result is not unexpected, because neglecting the third order derivatives in (5.1) implies that the effect of the quantum fluctuations, which is important well below threshold, is neglected. The above arguments allow us to give a quantitative condition for the validity of the truncated Wigner method in the model considered here. However, since the region below threshold is not accessible in the experiments due to the smallness of the emitted signal, the differences that appear between the exact and the approximate solution for $\delta < 0$ don't matter when comparing the theoretical and the experimental results on polariton correlations.

## VI. CONCLUSIONS

We have introduced a description of the statistics of the condensed polaritons in terms of a Wigner function. We have shown that in the truncated Wigner method [19] the evolution of Wigner function is described through a Fokker-Planck equation, which we solve analytically. The results for the correlations in the stationary state obtained in this framework are in astonishing good agreement with those obtained from the numerical solution of the master equation (2.5) and show a qualitative and, well above threshold, also quantitative [11] agreement with some experiments. Furthermore, in this description the contributions of the different noise effects that influence the polariton ground state statistics are explicitly defined and the calculations of the polariton statistics in the stationary state become very efficient. In the same framework, exploiting the equivalence between Fokker-Planck and Langevin descriptions of stochastic processes, we derive explicit expressions for the time dependent correlations of the polaritons close to the stationary state. We obtain expressions for the linewidth and for the relaxation rate of the polariton intensity in terms of the noise rates. The calculated values for these quantities are of the same order of magnitude of the measured ones. Finally, by solving the exact differential equation for the Wigner function, we are able to discuss the limits of validity of the truncated Wigner method in the present model. We show that above and around threshold the correlations calculated both with the exact and the approximate Wigner functions almost coincide, while below threshold strong deviations of the approximate correlations from the exact ones appear. Therefore, in the present model, the truncated Wigner method doesn't lead to correct results well below threshold. In fact in this regime the quantum fluctuations that are described by the third order derivatives in the exact equations and are disregarded in the truncated Wigner method become important. This last point may sheds also some light on the truncated Wigner method [19] currently applied in Bose-Einstein condensation

## ACKNOWLEDGEMENTS

We thank Y. Yamamoto and T. Horikiri for valued discussions and for having provided us with experimental datas presented in Fig. 5 prior to publication.

**APPENDIX**

As announced in Section II, in this Appendix we present some details on the derivation of the master equation (2.5) In order to simplify the notation we introduce the following definitions for the different terms in the polariton-polariton Hamiltonian (2.1) completed with the phonon-polariton interaction

$$H = H_1 + H_2$$
$$H_1 = \sum_{\mathbf{k}} \hbar \omega_{\mathbf{k}} P_{\mathbf{k}}^+ P_{\mathbf{k}} + W_{0,0,0} P_0^+ P_0^+ P_0 P_0$$
$$H_2 = \sum_{\mathbf{k},\mathbf{k}',\mathbf{q}} W_{\mathbf{k},\mathbf{k}'\mathbf{q}} (P_{\mathbf{q}}^+ P_{\mathbf{k}+\mathbf{k}'-\mathbf{q}}^+ P_{\mathbf{k}} P_{\mathbf{k}'} + h.c.) +$$
$$\sum_{\mathbf{q}} \hbar u \left(q_z^2 + q_{\parallel}^2\right)^{1/2} c_{q_z,\mathbf{q}_{\parallel}}^+ c_{q_z,\mathbf{q}_{\parallel}} + \sum_{\mathbf{k},\mathbf{k}',\mathbf{q}} H_{\mathbf{k},\mathbf{k}',(q_z,\mathbf{q}_{\parallel})} (c_{q_z,\mathbf{q}_{\parallel}} - c_{q_z,-\mathbf{q}_{\parallel}}^+) P_{\mathbf{k}}^+ P_{\mathbf{k}'} \delta_{\mathbf{k}',\mathbf{k}+\mathbf{q}_{\parallel}} \quad (A1)$$

where $c_{q_z,\mathbf{q}_{\parallel}}, c_{q_z,-\mathbf{q}_{\parallel}}^+$ are the creation and annihilation operators of the three-dimensional phonons respectively. The coefficients in the polariton-polariton interaction are defined as

$$W_{\mathbf{k},\mathbf{k}'\mathbf{q}} = \frac{1}{2}(V_{\mathbf{q},\mathbf{k}+\mathbf{k}'-\mathbf{q},\mathbf{k}} + V_{\mathbf{k}+\mathbf{k}'-\mathbf{q},\mathbf{q},\mathbf{k}})$$
$$V_{\mathbf{k},\mathbf{k}',\mathbf{q}} = \frac{6e^2 \lambda_X}{A\varepsilon} X_{\mathbf{k}} X_{\mathbf{k}'} X_{\mathbf{k}+\mathbf{q}} X_{\mathbf{k}'-\mathbf{q}} + \frac{\hbar\Omega_R 16\pi\lambda_X^2}{7A} X_{\mathbf{k}} X_{\mathbf{k}'-\mathbf{q}} \left( X_{\mathbf{k}'} |C_{\mathbf{k}+\mathbf{q}}| + |C_{\mathbf{k}'}| X_{\mathbf{k}+\mathbf{q}} \right)$$

Here $A$ is the quantization area, $\lambda_X$ is the exciton radius, $\Omega_R$ is the Rabi frequency and $X_\mathbf{k}, C_\mathbf{k}$ are the Hopfield coefficients for the exciton and the photon components of the polariton respectively.

The quantities that appear in the polariton-phonon interaction are defined as

$$H_{\mathbf{k},\mathbf{k}',(q_z,\mathbf{q}_\parallel)} = X_\mathbf{k} X_{\mathbf{k}'} \sqrt{\frac{\hbar \left(q_z^2 + q_\parallel^2\right)^{1/2}}{uV\rho_M}} \left(a_e I_e^\parallel(q_\parallel) I_e^\perp(q_z) - a_h I_h^\parallel(q_\parallel) I_h^\perp(q_z)\right)$$

$$I_{e,(h)}^\parallel(q_\parallel) = \left(1 + \frac{q_\parallel^2 \lambda_X^2 m_{e,(h)}^2}{4M^2}\right)^{-3/2} \quad \text{with } M = m_e + m_h$$

$$I_{e,h}^\perp(q_{z0}) = \frac{8\pi^2}{L_z q_\parallel (4\pi^2 - (L_z q_\parallel)^2)} \sin(L_z q_\parallel / 2)$$

Here, $u$ is the sound velocity, $\rho_M$ is the material density, $V$ the volume of the quantum well, $a_e$ and $a_h$ are the electron and hole deformation potentials respectively and $L_z$ is the width of the quantum well.

We start from the general equation for the reduced density matrix

$$\hbar \frac{d\rho}{dt} = -i[H,\rho] + \Lambda\rho \tag{A2}$$

Here $\Lambda$ is a damping operator such that

$$\Lambda\rho = \sum_{\mathbf{k}} \Lambda_{\mathbf{k}} = \sum_{\mathbf{k}} \gamma_{\mathbf{k}} \left( \left[ P_{\mathbf{k}}\rho, P_{\mathbf{k}}^{+} \right] + \left[ P_{\mathbf{k}}, \rho P_{\mathbf{k}}^{+} \right] \right) \quad (A3a)$$

$$\Lambda_{\mathbf{k}} P_{\mathbf{k}} = -\gamma_{\mathbf{k}} P_{\mathbf{k}} \quad (A3b)$$

$$\Lambda\rho(0) = 0 \quad (A3c)$$

As a first step, we integrate formally (*A2*) in time

$$\rho(t) = \exp\left[-i(L_1 + i\Lambda)t/\hbar\right]\rho(0) - i\int_0^t \exp\left[-i(L_1 + i\Lambda)\tau/\hbar\right] L_2 \rho(t-\tau) d\tau \quad (A4)$$

$$\rho(0) = \otimes_{q=0}^{K} |0\rangle_q {}_q\langle 0| \otimes \rho_{phon}^{eq}$$

$$L_1 Y = [\sum_{\mathbf{k}} \hbar\omega_{\mathbf{k}} P_{\mathbf{k}}^{+} P_{\mathbf{k}}, Y] + W_{0,0,0}\left[P_0^{+2} P_0^2, Y\right] = L_{11} Y + L_{12} Y$$

$$L_2 Y = [H_2, Y]$$

We insert (*A4*) into (*A1*), perform a Born approximation with respect to the term $L_{12}$ in (*A4*) and obtain

$$\hbar\frac{d\rho}{dt} = -i(L_{11} + L_{12} + i\Lambda)\rho - \int_0^t L_2 \exp\left[-i(L_{11} + i\Lambda)\tau/\hbar\right] L_2 \rho(t-\tau) d\tau +$$
$$L_2 \exp\left[-i(L_1 + i\Lambda)t/\hbar\right]\rho(0) \quad (A5a)$$

The master equation for $\rho_0$ is obtained from (*A5a*) by taking the trace over all wave vectors, reads

$$\hbar\frac{d\rho_0}{dt} = -i(L_{11} + L_{12} + i\Lambda)\rho_0 - Tr_{\{\mathbf{k}\neq 0\}}\int_0^t L_2 \exp\left[-i(L_{11} + i\Lambda)\tau/\hbar\right] L_2 \rho(t-\tau) d\tau \quad (A5b)$$

We take advantage of the relation

and rewrite (*A5b*) as

$$\hbar \frac{d\rho_0}{dt} = -i(L_{11} + L_{12} + i\Lambda)\rho_0 - Tr_{\{k\neq 0\}} \int_0^t L_2 L_2(\tau) \exp\left[-i(L_{11} + i\Lambda)\tau/\hbar\right]\rho(t-\tau)d\tau \quad (A7)$$

$L_2(\tau)Y = [H_2(\tau), Y]$ where $H_2(\tau)$ is the interaction Hamiltonian in which the time dependence of the operators is expressed through (A6).

We first consider the polariton-polariton interaction only. The polariton-phonon interaction contribution to the master equation is derived in a second step. This separation is justified through the fact that the influence of the polariton-phonon interaction on the dynamics of the ground state is negligible in the pump regime that we are considering.

We introduce a Markov approximation that allows us to perform the time integration by bringing the density operator out of the integral and letting the time go to infinity in the upper integration limit. We quote the result of these manipulations, by writing only the terms that contribute in the Boltzmann picture. After having introduced normal ordering they read for the interacting polariton part

$$\left[P_0 \sum_{\mathbf{k},\mathbf{k}'\neq 0} G_{\mathbf{k},\mathbf{k}',0} W^2_{\mathbf{k},\mathbf{k}',0} Tr_{\mathbf{k},\mathbf{k}'\neq 0}\left(P^+_{\mathbf{k}+\mathbf{k}'}P_{\mathbf{k}+\mathbf{k}'}\left(P^+_{\mathbf{k}}P_{\mathbf{k}}P^+_{\mathbf{k}'}P_{\mathbf{k}'} + P^+_{\mathbf{k}}P_{\mathbf{k}} + P^+_{\mathbf{k}'}P_{\mathbf{k}'} + 1\right)\right) P_0^+\right] + h.c. +$$

$$\left[P_0^+ \sum_{\mathbf{k},\mathbf{k}'\neq 0} G_{\mathbf{k},\mathbf{k}',0} W^2_{\mathbf{k},\mathbf{k}',0} Tr_{\mathbf{k},\mathbf{k}'\neq 0}\left((P^+_{\mathbf{k}+\mathbf{k}'}P_{\mathbf{k}+\mathbf{k}'} + 1)P^+_{\mathbf{k}}P_{\mathbf{k}}P^+_{\mathbf{k}'}P_{\mathbf{k}'}\right) P_0\right] + h.c. \quad (A8a)$$

and

$$\left[P_0^2, \sum_{\mathbf{k}\neq 0} G_{\mathbf{k},0} W^2_{\mathbf{k},-\mathbf{k},0} Tr_{\mathbf{k},-\mathbf{k}\neq 0}\left(P^+_{\mathbf{k}}P_{\mathbf{k}} + 1\right)\left(P^+_{-\mathbf{k}}P_{-\mathbf{k}} + 1\right)P_0^{+2}\right] + h.c. +$$

$$\left[P_0^{+2}, \sum_{\mathbf{k}\neq 0} G_{\mathbf{k},0} W^2_{\mathbf{k},-\mathbf{k},0} Tr_{\mathbf{k},-\mathbf{k}\neq 0}\left(P^+_{\mathbf{k}}P_{\mathbf{k}}\right)\left(P^+_{-\mathbf{k}}P_{-\mathbf{k}}\right)P_0^{+2}\right] + h.c. \quad (A8b)$$

$$G_{\mathbf{k},0} = \frac{2(2\gamma_0 + \gamma_{\mathbf{k}} + \gamma_{-\mathbf{k}'})}{(2\omega_0 - \omega_{\mathbf{k}} - \omega_{-\mathbf{k}})^2 + (2\gamma_0 + \gamma_{\mathbf{k}} + \gamma_{-\mathbf{k}})^2}$$

$$G_{\mathbf{k},\mathbf{k}',0} = \frac{2(\gamma_0 + \gamma_{\mathbf{k}} + \gamma_{\mathbf{k}'} + \gamma_{\mathbf{k}+\mathbf{k}'})}{(\omega_0 - \omega_{\mathbf{k}} - \omega_{\mathbf{k}'} + \omega_{\mathbf{k}+\mathbf{k}'})^2 + (\gamma_0 + \gamma_{\mathbf{k}} + \gamma_{\mathbf{k}'} + \gamma_{\mathbf{k}+\mathbf{k}'})^2}$$

These terms correspond to the ones that appear on the r.h.s. of equation (2.3) in Section II. The terms leading to frequency shifts are not written here. Introducing the Boltzmann factorization

$$Tr_{\{\mathbf{k}\}\neq 0}\left(P_{\mathbf{k}}^{+n}P_{\mathbf{k}}^{n}\rho\right)= \rho_0 \left(<P_{\mathbf{k}}^{+}P_{\mathbf{k}}>\right)^n \tag{A9a}$$

$$Tr_{\{\mathbf{k}\}\neq 0}\left(P_{\mathbf{k}}^{+m}P_{\mathbf{k}}^{n}\rho\right)= 0 \qquad m \neq n \tag{A9b}$$

into (A8b) we obtain an equation that formally resembles to the master equation

$$\hbar\frac{d}{dt}\rho_0(t) = -i\hbar\omega_0\left[P_0^+P_0,\rho_0(t)\right] - iW_{0,0,0}[P_0^+P_0^+P_0P_0,\rho_0(t)] + \Lambda_0\rho_0(t)) +$$
$$\Gamma_0\left(\left[P_0\rho_0(t),P_0^+\right]+h.c.\right) + \Delta_0\left(\left[P_0^+\rho_0(t),P_0\right]+h.c.\right) +$$
$$\Gamma_1\left(\left[P_0^2\rho_0(t),P_0^{+2}\right]+h.c.\right) + \Delta_1\left(\left[P_0^{+2}\rho_0(t),P_0^2\right]+h.c.\right) \tag{A10a}$$

$$\Gamma_0 = 2\operatorname{Re}\sum_{\mathbf{k},\mathbf{k}'\neq 0,}G_{\mathbf{k},\mathbf{k}',0}W_{\mathbf{k},\mathbf{k}',0}^{\phantom{0}2}\left[\left\langle P_{\mathbf{k}+\mathbf{k}'}^{+}P_{\mathbf{k}+\mathbf{k}'}\right\rangle\left(\left\langle P_{\mathbf{k}}^{+}P_{\mathbf{k}}\right\rangle+1\right)\left(\left\langle P_{\mathbf{k}'}^{+}P_{\mathbf{k}'( )}\right\rangle+1\right)\right] \tag{A10b}$$

$$\Delta_0 = 2\operatorname{Re}\sum_{\mathbf{k},\mathbf{k}'\neq 0,}G_{\mathbf{k},\mathbf{k}',0}W_{\mathbf{k},\mathbf{k}''0}^{\phantom{0}2}\left[\left(\left\langle P_{\mathbf{k}+\mathbf{k}-\mathbf{q}}^{+}P_{\mathbf{k}+\mathbf{k}'-\mathbf{q}}\right\rangle+1\right)\left\langle P_{\mathbf{k}}^{+}P_{\mathbf{k}}\right\rangle\left\langle P_{\mathbf{k}'}^{+}P_{\mathbf{k}'}\right\rangle\right] \tag{A10c}$$

$$\Gamma_1 = 2\operatorname{Re}\sum_{\mathbf{k}\neq 0}G_{\mathbf{k},0}W_{\mathbf{k},-\mathbf{k},0}^{\phantom{0}2}\left(\left(<P_{\mathbf{k}}^{+}P_{.\mathbf{k}}>+1\right)\left(<P_{-\mathbf{k}}^{+}P_{-\mathbf{k}}>+1\right)\right) \tag{A10d}$$

$$\Delta_1 = 2\operatorname{Re}\sum_{\mathbf{k}\neq 0}G_{\mathbf{k}}W_{,\mathbf{k},-\mathbf{k},0}^{\phantom{0}2}\left(<P_{\mathbf{k}}^{+}P_{.\mathbf{k}}><P_{-\mathbf{k}}^{+}P_{-\mathbf{k}}>\right) \tag{A10e}$$

Notice that here the rates that multiply the commutators in general depend on the dynamically determined populations of the polariton modes with $\mathbf{k}\neq 0$.

The polariton-phonon interaction is now introduced using the same considerations leading from (A7) to (A9). However, in this case the phonons may be considered as a reservoir whose temperature is given by the lattice temperature. Therefore, we don't need to introduce the Boltzmann factorization for the expectation values involving phonon operators only. Using the standard approach for deriving the dynamics of the polaritons up to second order in the polariton-phonon interaction, the contribution to the general master equation is given below where the trace over the reservoir operators has already been performed

$$\left[ P_0 \sum_{\mathbf{k},\mathbf{k'},\mathbf{q}\neq 0} G^{ph}_{(\mathbf{q},\mathbf{q'},\mathbf{k})} W^{ph2}_{\mathbf{q}\to\mathbf{q'},\mathbf{k}} Tr_{\mathbf{k},\mathbf{k'},\mathbf{q}\neq 0} \left( P^+_{\mathbf{k}} P_{\mathbf{k}} + 1 \right), P^+_0 \right] + h.c. +$$

$$\left[ P^+_0 \sum_{\mathbf{k},\mathbf{k'}\neq 0} G^{ph}_{(\mathbf{q},\mathbf{q'},\mathbf{k})} W^{ph2}_{\mathbf{q}\to\mathbf{q'},\mathbf{k}} Tr_{\mathbf{k},\mathbf{k'},\mathbf{q}\neq 0} P^+_{\mathbf{k}} P_{\mathbf{k}}, P_0 \right] + h.c. \qquad (A11a)$$

$$G^{ph}_{(\mathbf{q},\mathbf{q'},\mathbf{k})} = \frac{i\left(E^{ph}_{\mathbf{k}} - |\hbar\omega_{\mathbf{q}} - \hbar\omega_{\mathbf{q'}}|\right) + \left(\gamma_{\mathbf{q}} + \gamma_{\mathbf{q'}}\right)}{\left(E^{ph}_{\mathbf{k}} - |\hbar\omega_{\mathbf{q}} - \hbar\omega_{\mathbf{q'}}|\right)^2 + \left(\gamma_{\mathbf{q}} + \gamma_{\mathbf{q'}}\right)^2} \qquad (A11b)$$

$$W^{ph}_{\mathbf{q}\to\mathbf{q'},\mathbf{k}} = \frac{2\pi}{\hbar} \delta_{\mathbf{q}-\mathbf{q'}-\mathbf{k}_{\parallel}} \left| X^*_{\mathbf{q'}} \langle \mathbf{q'} | \langle 0_{\mathbf{k}} | H_{exc-ph} | 1_{\mathbf{k}} \rangle | \mathbf{q} \rangle X_{\mathbf{q}} \right|^2 \left[ n_B\left(E^{ph}_{\mathbf{k}}\right) + \theta(\omega_{\mathbf{q}} - \omega_{\mathbf{q'}}) \right] \qquad (A11c)$$

where $H_{exc-ph}$ is the deformation potential interaction, $|n_{\mathbf{k}}\rangle$ is a phonon number state, $|\mathbf{q}\rangle$ is the exciton state of wave vector $\mathbf{q}$, $E^{ph}_{\mathbf{k}}$ is the energy of a phonon with wave vector $\mathbf{k}$, $n_B\left(E^{ph}_{\mathbf{k}}\right)$ is the Bose occupation function and $\theta(\omega_{\mathbf{q}} - \omega_{\mathbf{q'}})$ is the Heavyside function. Notice that here the sums span over the whole lower polariton branch. The terms (*A11*) contribute to the relaxation and injection rates into the ground state.

We notice that from (*A5a*) we can derive the equations describing the dynamics of the polariton occupations with wave vector $\mathbf{k}\neq 0$. We have to multiply (*A5*) by the polariton number operator $P_{\mathbf{k}} P_{\mathbf{k}}$, take the trace over all modes and perform the manipulations (*A8*) and (*A9*). We obtain

$$\hbar \frac{d}{dt}\langle P_\mathbf{k}^\dagger P_\mathbf{k}\rangle = -2(\gamma_\mathbf{k} + \Gamma_{\mathbf{k},TOT} - \Delta_{\mathbf{k},TOT})\langle P_\mathbf{k}^\dagger P_\mathbf{k}\rangle + 2\Delta_{\mathbf{k},TOT} +$$

$$W_{0,\mathbf{k},-\mathbf{k}}^2 \operatorname{Re}\left\{\frac{2\langle P_0^\dagger P_0\rangle^2 (\langle P_\mathbf{k}^\dagger P_\mathbf{k}\rangle + \langle P_{-\mathbf{k}}^\dagger P_{-\mathbf{k}}\rangle + 1)}{[i\hbar(\omega_0 - \omega_\mathbf{k}) + \gamma_\mathbf{k} + \gamma_0]}\right\} -$$

$$W_{0,\mathbf{k},-\mathbf{k}}^2 \operatorname{Re}\left\{\frac{2(\langle P_0^\dagger P_0\rangle + 1)(\langle P_\mathbf{k}^\dagger P_\mathbf{k}\rangle \langle P_{-\mathbf{k}}^\dagger P_{-\mathbf{k}}\rangle + 1)}{[i\hbar(\omega_0 - \omega_\mathbf{k}) + \gamma_\mathbf{k} + \gamma_0]}\right\} + F_\mathbf{q} \quad (A12)$$

In (A12), $\gamma_\mathbf{k}$ is the radiative linewidth, $\omega_\mathbf{k}$ is the lower polariton dispersion and

$$\Gamma_{\mathbf{q},TOT} \equiv \Gamma_\mathbf{q}^{pp} + \Gamma_\mathbf{q}^{ph}, \quad (A13a)$$

$$\Gamma_\mathbf{q}^{pp} = 2\sum_{\mathbf{k},\mathbf{k}'} \operatorname{Re} G_{(\mathbf{k},\mathbf{k}',\mathbf{k}+\mathbf{k}'-\mathbf{q})} W_{\mathbf{k},\mathbf{k}',\mathbf{q}}^2 (N_\mathbf{k}+1)(N_{\mathbf{k}'}+1) N_{\mathbf{k}+\mathbf{k}'-\mathbf{q}},$$

$$\Gamma_\mathbf{q}^{ph} = 2\sum_{\mathbf{q}',\mathbf{k}} \operatorname{Re} G_{(\mathbf{q},\mathbf{q}',\mathbf{k})}^{ph} W_{\mathbf{q}\to\mathbf{q}',\mathbf{k}}^{ph} (N_{\mathbf{q}'}+1)$$

$$\Delta_{\mathbf{q},TOT} \equiv \Delta_\mathbf{q}^{pp} + \Delta_\mathbf{q}^{ph}, \quad (A13b)$$

$$\Delta_\mathbf{q}^{pp} = 2\sum_{\mathbf{k},\mathbf{k}'} \operatorname{Re} G_{(\mathbf{k},\mathbf{k}',\mathbf{k}+\mathbf{k}'-\mathbf{q})} W_{\mathbf{k},\mathbf{k}',\mathbf{q}}^2 (N_\mathbf{k}+1)(N_{\mathbf{k}'}+1) N_{\mathbf{k}+\mathbf{k}'-\mathbf{q}},$$

$$\Delta_\mathbf{q}^{ph} = 2\sum_{\mathbf{q}',\mathbf{k}} \operatorname{Re} G_{(\mathbf{q},\mathbf{q}',\mathbf{k})}^{ph} W_{\mathbf{q}'\to\mathbf{q},\mathbf{k}}^{ph} N_{\mathbf{q}'}$$

$$G_{(\mathbf{k},\mathbf{k}',\mathbf{k}+\mathbf{k}'-\mathbf{q})} = \frac{i\hbar(\omega_\mathbf{q} + \omega_{\mathbf{k}+\mathbf{k}'-\mathbf{q}} - \omega_\mathbf{k} - \omega_{\mathbf{k}'}) + (\gamma_\mathbf{q} + \gamma_{\mathbf{k}+\mathbf{k}'-\mathbf{q}} + \gamma_\mathbf{k} + \gamma_{\mathbf{k}'})}{\hbar^2(\omega_\mathbf{q} + \omega_{\mathbf{k}+\mathbf{k}'-\mathbf{q}} - \omega_\mathbf{k} - \omega_{\mathbf{k}'})^2 + (\gamma_\mathbf{q} + \gamma_{\mathbf{k}+\mathbf{k}'-\mathbf{q}} + \gamma_\mathbf{k} + \gamma_{\mathbf{k}'})^2},$$

$$G_{(\mathbf{q},\mathbf{q}',\mathbf{k})}^{ph} = \frac{i(E_\mathbf{k}^{ph} - |\hbar\omega_\mathbf{q} - \hbar\omega_{\mathbf{q}'}|) + (\gamma_\mathbf{q} + \gamma_{\mathbf{q}'})}{(E_\mathbf{k}^{ph} - |\hbar\omega_\mathbf{q} - \hbar\omega_{\mathbf{q}'}|)^2 + (\gamma_\mathbf{q} + \gamma_{\mathbf{q}'})^2}.$$

Notice that the form (A5) also a master equation for the density operator of any mode with a wave vector $\mathbf{k} \neq 0$ can be derived. In fact, it is sufficient to put into evidence in (A5b) the polariton with wave vector $\mathbf{k} \neq 0$ instead of the ones with $\mathbf{k} = 0$ all other steps remaining the same as the ones leading to (A12). Along this same lines we obtain the equivalent of (A13) for any value of $\mathbf{k} \neq 0$.

**FIGURE CAPTIONS**

Fig. 1. Approximate Wigner probability distribution $W(\alpha,\alpha^*)$ as a function of $|\alpha|$ for different normalized pump intensities close to the threshold and with a detuning $D = 7$ $meV$. Solid line $F/F_s = 1$. Broken line $F/F_s = 1.5$. Material parameters appropriate to GaAs are used.

Fig. 2. Stationary normalized second order correlation function $g^{(2)}(0)$ as a function of the normalized pump intensity. Solid line: result obtained with the approximate Wigner function (2.14). Broken line: results obtained from the master equation (2.5). Dots: experimental points [10]. The vertical line indicates the position of the threshold. Material parameters appropriate to CdTe are used.

Fig. 3. Stationary normalized second order correlation function $g^{(2)}(0)$ as a function of the normalized pump intensity evaluated with the approximate Wigner function (2.14) for two different detunings $D$. Broken line $D = 2$ $meV$. Solid line $D = 7$ $meV$. The insert indicates that a reduced minimum is also present for $D= 2$ $meV$. The vertical line indicates the position of the threshold. Material parameters appropriate to GaAs are used.

Fig. 4. Approximate Wigner probability distribution as a function of $|\alpha|$ for different normalized pump intensities and detuning. Grey solid line $F/F_s = 1$, $D = 7$ $meV$. Black solid line $F/F_s = 2$, $D = 7$ $meV$. Grey broken line $F/F_s = 1$, $D = 2$ $meV$. Black broken line $F/F_s = 1.5$, $D = 2$ $meV$. Material parameters appropriate to GaAs are used.

Fig. 5. Stationary normalized second order correlation function $g^{(2)}(0)$ as a function of the normalized pump intensity. Dots: experimental results[11]. Solid line: result obtained with the approximate Wigner function (2.14) and with material parameters appropriate to GaAs. The vertical line indicates the position of the threshold.

Fig. 6. Modulus of the amplitude as a function of the normalized pump intensity. Solid line: result obtained with the approximate Wigner function (2.14). Broken line: result obtained from the master equation (2.5). The vertical line indicates the position of the threshold. The difference between the two approaches is relevant only for high intensities as visualized in the insert. Material parameters appropriate to CdTe are used.

Fig. 7. Solid line: plot of the linewidth $\gamma_c$ (Eq. 4.9b) as a function of the normalized pump intensity. Broken line: plot, as a function of the normalized pump intensity, of the parametric diffusion $(\Gamma_1 + \Delta_1)$ (the sum of injection and dissipation rates), which dominates $\gamma_c$ for high pump intensities. The vertical line indicates the position of the threshold. Material parameters appropriate to CdTe are used.

Fig. 8. Solid line: plot of the relaxation rate of the polaritons (see Eq. 4.14) as a function of the normalized pump intensity. Broken line: plot of the relaxation rate where of the parametric diffusion $(\Gamma_1 + \Delta_1)$ (the sum of injection and dissipation rates) is neglected. The insert indicates the decreasing of the relaxation rate below threshold. The vertical line indicates the position of the threshold. Material parameters appropriate to CdTe are used.

**Fig. 1**

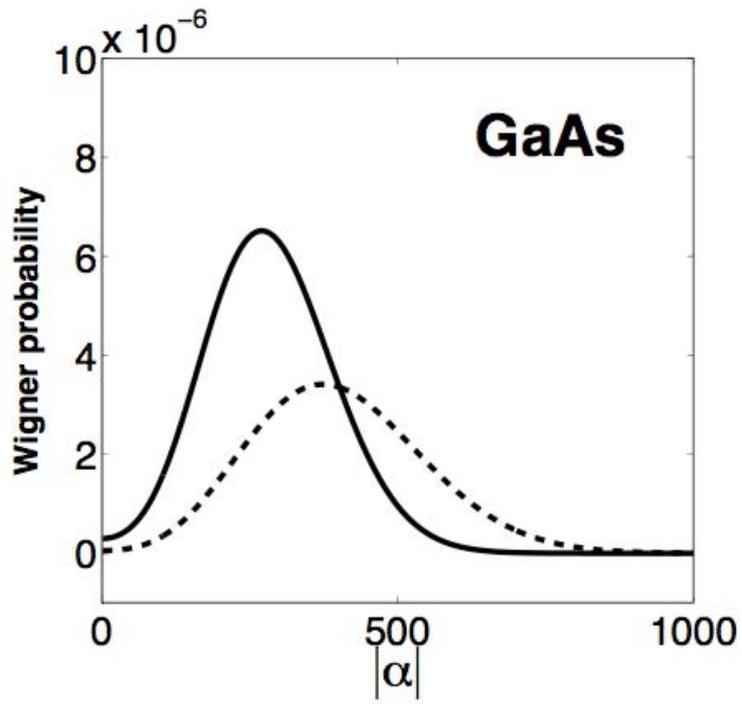

**Fig. 2**

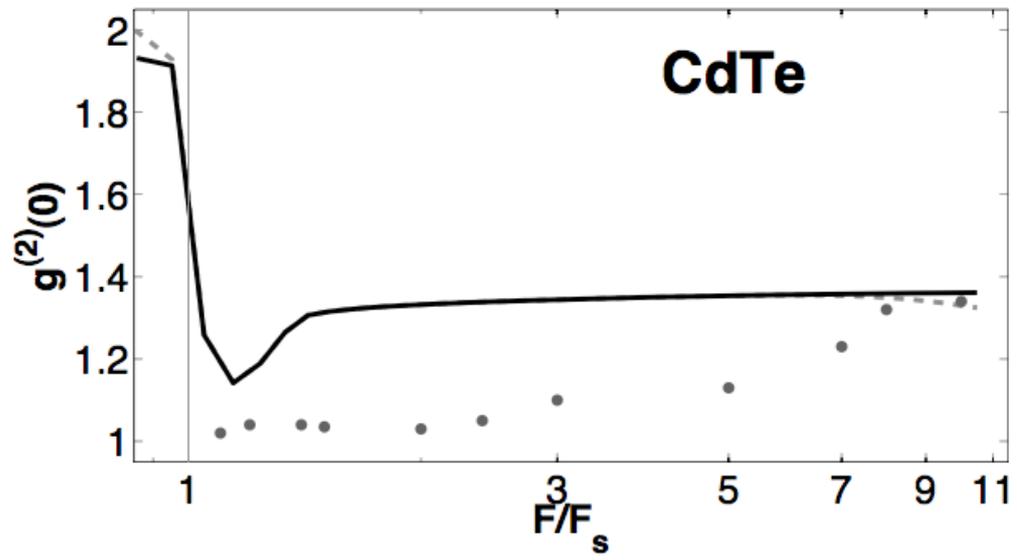

**Fig. 3**

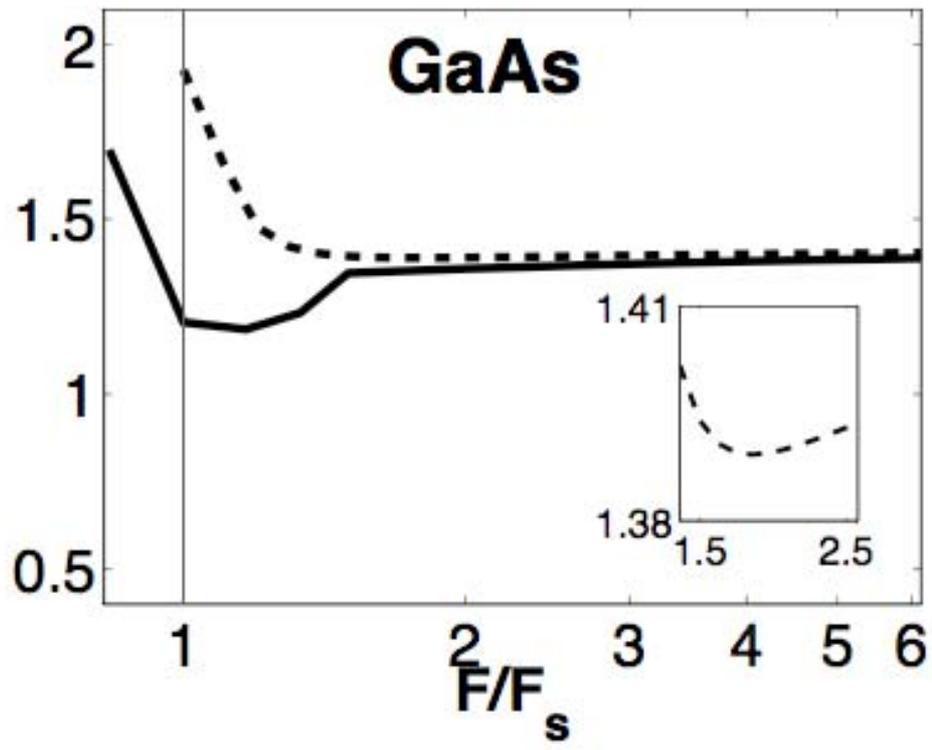

**Fig. 4**

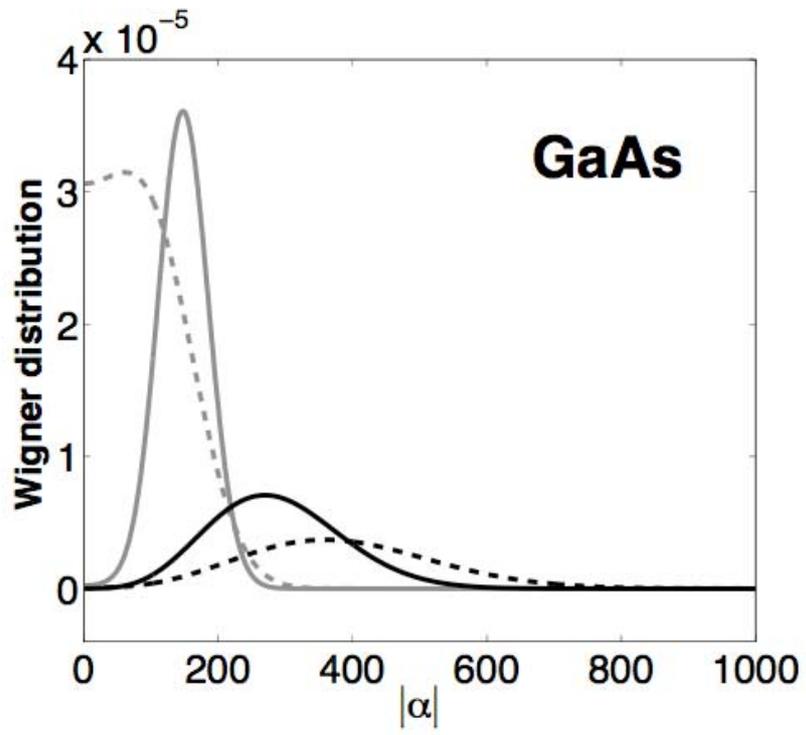

**Fig. 5**

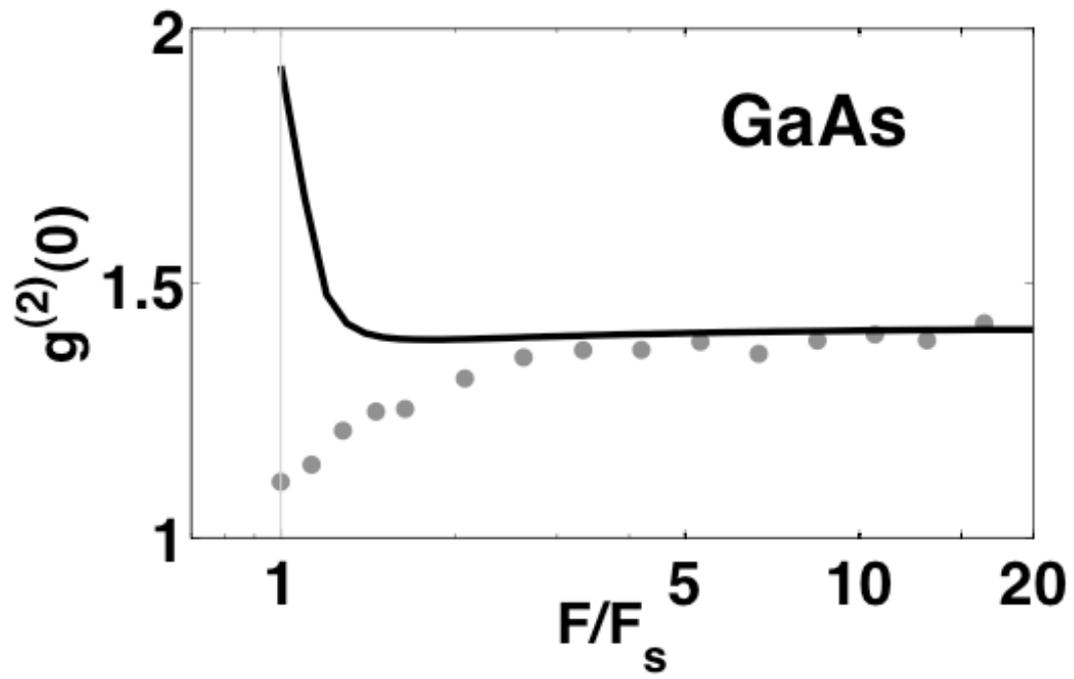

**Fig. 6**

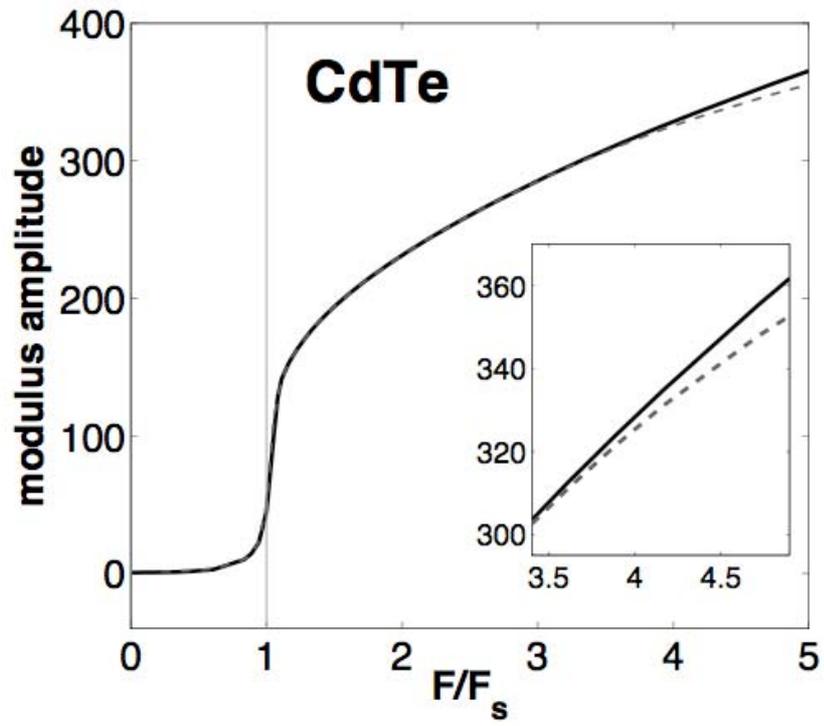

**Fig. 7**

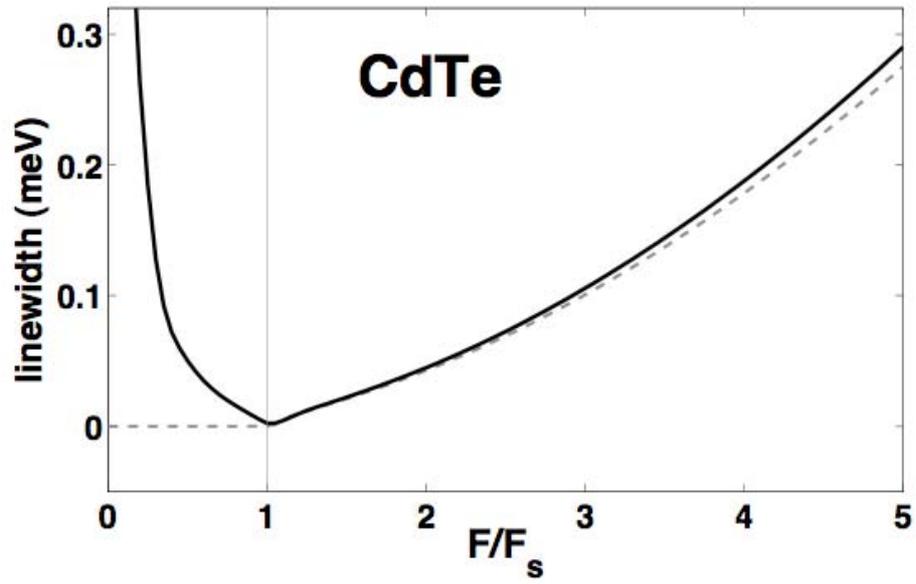

**Fig. 8**

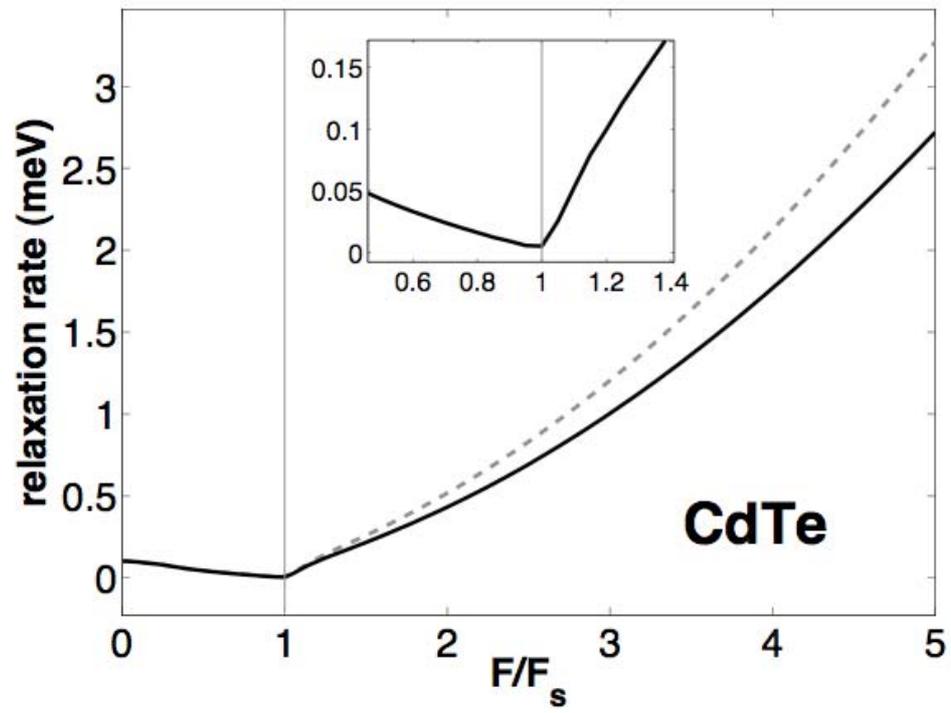